\documentclass[aoas,MSNbibl,nameyear,dvips]{arximspdf}
\usepackage{graphicx}
\doi{10.1214/12-AOAS547} 
\volume{6}
\issue{3}
\pubyear{2012}
\firstpage{895}
\lastpage{927}

\makeatletter

\newcommand{\real}{\mathbb{R}}
\newcommand{\err}{\varepsilon}
\newcommand{\eps}{\epsilon}

\newcommand{\wh}{\widehat}
\newcommand{\wt}{\widetilde}
\newcommand{\tran}{\mathsf{T}}

\newcommand{\sumdot}{{\mbox{\tiny$\bullet$}}}

\newcommand{\bsi}{\mathbf{i}}
\newcommand{\bsip}{\mathbf{i}'}
\newcommand{\bsipp}{\mathbf{i}''}
\newcommand{\bsippp}{\mathbf{i}'''}

\newcommand{\one}{\mathbf{1}}

\newcommand{\wbsi}{W_{\bsi}}
\newcommand{\xbsi}{X_{\bsi}}
\newcommand{\ybsi}{Y_{\bsi}}
\newcommand{\zbsi}{Z_{\bsi}}

\newcommand{\wbsip}{W_{\bsip}}
\newcommand{\xbsip}{X_{\bsip}}
\newcommand{\ybsip}{Y_{\bsip}}
\newcommand{\zbsip}{Z_{\bsip}}

\newcommand{\xbsipp}{X_{\bsipp}}
\newcommand{\zbsipp}{Z_{\bsipp}}

\newcommand{\xbsippp}{X_{\bsippp}}
\newcommand{\zbsippp}{Z_{\bsippp}}

\newcommand{\wbsib}{W_{\bsi,b}}
\newcommand{\wbsipb}{W_{\bsip,b}}

\newcommand{\nbsiu}{N_{{\bsi},u}}
\newcommand{\nbsipu}{N_{{\bsip},u}}
\newcommand{\nbsippu}{N_{{\bsipp},u}}

\newcommand{\nbsik}{N_{{\bsi},k}}

\newcommand{\nubsiu}{\nu_{{\bsi},u}}
\newcommand{\nubsipu}{\nu_{{\bsip},u}}
\newcommand{\nubsippu}{\nu_{{\bsipp},u}}

\newcommand{\nusigbar}{\overline{\nu_u\sigma^2_{u}}}
\newcommand{\nukbar}{\overline{\nu}_k}

\newcommand{\nusetj}{\nu_{\{j\}}}
\newcommand{\gamsetj}{\gamma_{\{j\}}}
\newcommand{\gamset}[1]{\gamma_{#1}}
\newcommand{\nuset}[1]{\nu_{#1}}
\newcommand{\sigsqsetj}{\sigma^2_{\{j\}}}

\newcommand{\sbsi}{\sum_{\bsi}}
\newcommand{\sbsip}{\sum_{\bsip}}
\newcommand{\sbsipp}{\sum_{\bsipp}}
\newcommand{\sbsippp}{\sum_{\bsippp}}

\newcommand{\sneu}{\sum_{u\ne\varnothing}}
\newcommand{\sneup}{\sum_{u'\ne\varnothing}}

\newcommand{\dbin}{\operatorname{Bin}}
\newcommand{\dexp}{\operatorname{Exp}}
\newcommand{\dpoi}{\operatorname{Poi}}
\newcommand{\dustd}{\mathbf{U}}

\newcommand{\miip}{M_{\bsi\bsip}}

\newcommand{\oiuipu}{\one_{\bsi_u=\bsip_u}}

\newcommand{\e}{\mathbb{E}}
\newcommand{\var}{\operatorname{Var}}
\newcommand{\cov}{\operatorname{Cov}}

\newcommand{\nb}{\mathrm{NB}}
\newcommand{\nbb}{\mathrm{NBB}}
\newcommand{\re}{\mathrm{RE}}
\newcommand{\bpb}{\mathrm{PW}}

\newcommand{\nbbvar}{\wt\var_{\nbb}}

\newtheorem{theorem}{Theorem}
\newtheorem{lemma}{Lemma}

\makeatother

\begin{document}
\begin{frontmatter}

\title{Bootstrapping data arrays of arbitrary order\thanksref{T1}}
\runtitle{Bootstrapping data arrays}

\thankstext{T1}{Supported by NSF Grant DMS-09-06056 and by Nokia.}

\begin{aug}
\author[A]{\fnms{Art B.} \snm{Owen}\corref{}\ead[label=e1]{owen@stat.stanford.edu}}
\and
\author[B]{\fnms{Dean} \snm{Eckles}}

\runauthor{A. B. Owen and D. Eckles}
\affiliation{Stanford University, and Stanford University and
Facebook, Inc.}
\address[A]{Department of Statistics\\
Stanford University\\
Sequoia Hall\\
Stanford, California 94305\\
USA\\
\printead{e1}} 
\address[B]{Facebook Inc.\\
1601 Willow Road\\
Menlo Park, California 94025\\
USA}
\end{aug}

\received{\smonth{6} \syear{2011}}
\revised{\smonth{12} \syear{2011}}

%
\begin{abstract}
In this paper we study a bootstrap strategy for estimating the variance
of a mean taken over large multifactor crossed random effects data
sets. We apply bootstrap reweighting independently to the levels of
each factor, giving each observation the product of independently
sampled factor weights. No exact bootstrap exists for this problem
[McCullagh (2000) \textit{Bernoulli} \textbf{6}
285--301]. We show that the proposed bootstrap is mildly
conservative, meaning biased toward overestimating the variance, under
sufficient conditions that allow very unbalanced and heteroscedastic
inputs. Earlier results for a resampling bootstrap only apply to two
factors and use multinomial weights that are poorly suited to online
computation. The proposed reweighting approach can be implemented in
parallel and online settings. The results for this method apply to any
number of factors. The method is illustrated using a $3$ factor data
set of comment lengths from Facebook.
\end{abstract}

%

\end{frontmatter}

\section{Introduction}

Large sparse data sets with two or more crossed
random effects commonly arise from electronic commerce
and Internet services, and we may expect them
to arise in other settings as automated data
gathering becomes more prevalent.
Such data often have no IID
structure for us to draw on.
For example, with the famous Netflix data
[\citet{bennlann2007}]
multiple ratings from the same
viewer are dependent.
Similarly, multiple ratings on the same movie
are dependent.
Neither rows nor columns are IID,
and a crossed random effects model with
interactions is a~more reasonable structure.

In Internet data
there can easily be more than two
crossed factors. The individual factors
could be user account numbers, IP
addresses, URLs, search query strings or
identifiers for documents placed in web pages.
The response variable might be a measure of user
engagement such\vadjust{\goodbreak} as time spent reading,
or system performance such as the load times for pages under
different versions of software.

The crossed random effects setting is
challenging for inference.
Methods in \citet{searcasemccu1992}
rely on Gaussian data assumptions
and outside of balanced cases,
the necessary linear algebra becomes prohibitively expensive
on large problems.

We might therefore turn to resampling. For
IID data, the bootstrap provides
reliable variance estimates and confidence
intervals under very weak assumptions
on the mechanism generating our data.
But \citet{mccu2000} proved that there does
not exist an exact bootstrap algorithm for
crossed random effects. Specifically,
if $X_{ij}=\mu+a_i+b_j+\err_{ij}$
for independent\vspace*{1pt} mean~$0$ random
variables $a_i$, $b_j$ and $\err_{ij}$
with variances $\sigma^2_A$, $\sigma^2_B$ and
$\sigma^2_E$, respectively, then no resampling
method, from a very broad class, will provide
an unbiased estimate of
$\var((IJ)^{-1}\sum_{i=1}^I\sum_{j=1}^JX_{ij})$.\vspace*{1pt}

One approach to bootstrapping crossed data
is to independently bootstrap the indices of each factor.
In bootstrapping a factor we are putting a~random
multinomially distributed weight on the levels of that factor.
For an $r$-fold data set, the observation $X_{i_1i_2\cdots i_r}$
gets a weight
$W^{*b}_{i_1i_2\cdots i_r} = \prod_{j=1}^r W_{j,i_j}^{*b}$,
where~$W_{j,i_j}^{*b}$ is\vspace*{1pt} the weight on level $i_j$
of the $j$th factor in the $b$th bootstrap reweighting.
For each $j$ and each $b$,
weight vectors $(W^{*b}_{j,1},W^{*b}_{j,2},\ldots,W^{*b}_{j,N_j})$~are
sampled independently.
Given weights on all the data, we compute a~weighted version
of the statistic(s) of interest to get the $b$th
bootstrap value.\looseness=-1

Bootstrapping with a product of multinomial
weights has been studied before, for $r=2$.
\citet{brenharrhans1987} and \citet{wile2001}
use it to study variance components in educational test data.
\citet{mccu2000} shows that independently bootstrapping
the rows and columns of a data matrix gives a
mildly conservative estimate of variance. That is,
it has a~positive bias that is usually relatively small.
\citet{mccu2000} considered balanced crossed
random effects (no missing values) with homoscedastic variance components.
\citet{pbs}
shows that this bootstrap remains conservative
(and usually mildly so)
for sparsely sampled unbalanced crossed random effects
allowing for heteroscedasticity.
That framework allows every row and
column (e.g., customer and movie) and
even every interaction to have its own variance.
Resampling is then reliable and it spares
the analyst from having to estimate all of those variances.

The random weighting that we favor is a product
of completely independent weights: $W_{i_1i_2,\ldots,i_r}^{*b}
= \prod_{j=1}^r W_{ji_j}^{*b}$, where
for each $b$ and each $j$, $W^{*b}_{ji_j}$ are~IID
weights with mean $1$ and variance $1$.
For these large data sets, methods that reweight data
via IID random weights
[\citet{rubi1981}, \citet{newtraft1994}]
are an appealing alternative to
the multinomial weights used in resampling.
First, it is
simpler to apply independent reweighting to large
scale parallelized computations, as
is done in online bagging and boosting
[\citet{oza2001}, \citet{leeclyd2004}].\vadjust{\goodbreak}
The reason is that large data sets are
stored in a distributed fashion and
then multinomial sampling brings substantial
communication and synchronization costs.
Second, resampling simplifies variance
expressions by avoiding the negative dependence
from the multinomial distribution.
This makes it easier to develop expressions for
problems with more than two factors.

Using notation and approximations defined below, the main facts are as
follows. For $r=2$ factors, we suppose the data are sampled by a random
effects model with variance components $\sigma^2_{\{1\}}$,
$\sigma^2_{\{2\}}$ and $\sigma^2_{\{1,2\}}$ corresponding\vspace*{1pt}
to the main effects and interaction, respectively. We can express the
variance of the sample average of $N$ observations in the form
$(\nu_{\{1\}}\sigma^2_{\{1\}}+\nu_{\{2\}}\sigma^2_{\{2\}}+\sigma ^2_{\{
1,2\}})/N$. The subscripted $\nu$ quantities are easily computable from
the data and we give explicit formulas. Naive bootstrapping produces an
estimate close to
$(\sigma^2_{\{1\}}+\sigma^2_{\{2\}}+\sigma^2_{\{1,2\}})/N$\vspace*{1pt}
which is grossly inadequate because it turns out that often
$\nu_{\{j\}}\gg1$.
For instance, in the Netflix data set,
the largest
$\nu_{\{j\}}$ is about $56\mbox{,}200$.

Resampling both rows and columns leads to a variance
estimate close to
$((\nu_{\{1\}}+2)\sigma^2_{\{1\}}+(\nu_{\{2\}}+2)\sigma^2_{\{2\}
}+3\sigma^2_{\{1,2\}})/N$,
which is mildly conservative when $\nu_{\{j\}}\gg1$
and the $\sigma$'s are of comparable magnitude.
It is up to three times as large as it
should be in the event that $\sigma^2_{\{j\}}\ll\sigma^2_{\{1,2\}}$.
Being conservative by a factor of at most $3$ is
far more acceptable than underestimating variance
by as much as $56\mbox{,}200$.

Our main contributions are as follows:
\begin{longlist}[(3)]
\item[(1)] We show that a naive bootstrap suitable for IID
settings severely underestimates the
variance of the sample mean, when $r=2$, while the
product strategy mildly overestimates it.
These facts were known for resampling, but we show it
also for reweighting.

\item[(2)] We generalize\vspace*{1pt} the
reweighting results to $r\ge2$ factors.
In particular, for the homoscedastic setting,
the $3\sigma^2_{\{1,2\}}$ variance
contribution from the case $r=2$
becomes $(2^r-1)\sigma^2_{\{1,2,\ldots,r\}}$.
We find\vspace*{1pt} expressions for all $2^r-1$ variance coefficients.
Under reasonable conditions, for which we note
exceptions, this bootstrap magnifies
a $k$-factor variance component by roughly $2^k-1$.
Under simply described conditions,
the $k=1$ terms dominate the variance and
then the variance magnification becomes negligible.

\item[(3)]
We introduce a heteroscedastic random effects
model in which every nonempty subset of factors
contributes a random effect.
The product weighted bootstrap remains mildly conservative
even when every factorial effect
for every observation has a distinct variance, so long
as all the variances are uniformly bounded away
from $0$ and infinity.
\end{longlist}

An outline of the paper is as follows.
Section~\ref{secnotation} introduces
our notation for the random effects model
and some observation counts and then defines
the random\vadjust{\goodbreak} effects variance that we seek to estimate.
Section~\ref{secnaive} considers naive bootstrap
methods that simply resample or reweight
the observations as if they were IID.
They seriously underestimate the true
variance unless the only nonzero variance
component is that of the highest order interaction.
Reweighting has a slight advantage because
it allows one to step up the sampling variance
to compensate for cases where the naive
bootstrap variance
is only a modest underestimate.
Section~\ref{secproposal} introduces a
factorial reweighting bootstrap strategy.
For data with $r=2$ factors, the
reweighting results closely match
the resampling results from \citet{pbs}.
This section includes an interpretable approximation
to the exact bootstrap variance.
Section~\ref{sechet}
considers the heteroscedastic case, where
every variance component at every
combination of its factors
has its own variance parameter.
When the main effects are dominant, then
the proposed bootstrap closely matches
the desired variance even in the heteroscedastic setting.
Section~\ref{secnesting} describes repeated observations
and factors nested inside the ones being reweighted.
Section~\ref{secexample}
has a numerical example from Facebook.
In that data set,
UK-based users make longer comments
than do US-based users, when posting from mobile
devices. The reverse holds for comments made
at Facebook's standard web interface.
The differences are small, but statistically significant,
even after taking account of a three factor structure
(commenter, sharer and URL).
The proofs appear in the \hyperref[app]{Appendix}.

Although the product reweighting algorithm is
simple, its analysis in the random effects
context is very technical.
Section~\ref{secdiscussion} discusses
some larger statistical issues.
Among these are the reasons that we do not model the possible
informativeness of the missing data mechanism,
the reasons for focusing on the bootstrap variance of a sample mean,
and the motivation for considering the heteroscedastic
random effects model, which contains many
more parameters than observations.

\section{Notation and random effects model}\label{secnotation}

The random variables of interest take the form
$X_{i_1,i_2,\ldots,i_r}\in\real^d$ for
integers $i_j\ge1$ and $j=1,\ldots,r$.
To simplify notation, we write $\xbsi$
for $\bsi=(i_1,\ldots,i_r)$.
We work with $X$ of dimension $d=1$. The generalization
to $d\ge1$ is straightforward.
We have in mind applications where each value of
$i_j$ corresponds to one level of a categorical
variable with many potential values.
In Internet applications,
index values $i_j$ might represent users, URLs, IP addresses,
ads, query strings and so on. There may be no
a priori upper bound on the number of distinct
levels for $i_j$.

The data are composed of $N$ of these
random variables, where $1\le N<\infty$.
The binary variable $\zbsi$ takes the value $1$
when observation $\xbsi$ is present
and $\zbsi=0$ when $\xbsi$ is absent.
We work conditionally on $\zbsi$
so that they are nonrandom.
In practice, the pattern of missingness
in $\zbsi$ may be important.
As with prior work,
we avoid modeling $\zbsi$ in order
to focus on estimating variance,
apart from some brief remarks
in Section~\ref{secdiscussion}.

The letters $u$ and $v$ denote subsets of $[r]\equiv\{1,\ldots,r\}$
throughout. The summation $\sum_u$ is taken over all $2^r$ subsets of\vadjust{\goodbreak}
$[r]$, and other summations,\vspace*{1pt} such as $\sum_{v\supseteq
u}$, denote sums over the first named set (here $v$) subject to the
indicated condition with the other set(s) (here $u$) held fixed. The
index $\bsi_u$ extracts the components $i_j$ for $j\in u$. Then
$\bsi_u=\bsip_u$ means that $i_j=i'_j$ for all $j\in u$.

Our \textit{$r$-fold crossed random effects model}
is
%
%
\begin{equation}\label{eqreff}
X_{\bsi}
=\mu+ \sum_{u\ne\varnothing} \err_{\bsi,u},
\end{equation}
where $\mu\in\real$ and $\err_{\bsi,u}$ are mean $0$ random variables
that depend on $\bsi$ only through~$\bsi_u$. We have
$\err_{\bsi,u}=\err_{\bsip,u}$ if $\bsi_u=\bsip_u$ and $\err_{\bsi,u}$
independent of $\err_{\bsip,u}$ otherwise. The covariance of
$\err_{\bsi,u}$ and $\err_{\bsi',u'}$ is
%
%
\begin{equation}\label{eqcovreff}
\cov(\err_{\bsi,u},\err_{\bsi',u'})
=\e(\err_{\bsi,u}\err_{\bsi',u'})=
\sigma^2_u\one_{u=u'}\oiuipu
\end{equation}
for $\sigma^2_u<\infty$.

To illustrate the model notation, suppose that $r=2$
and one observation is at $\bsi= (38,44)$
while another is at $\bsip= (38,19)$.
Then
%
%
\begin{eqnarray}\label{eqexample38}
X_{\bsi}&=&X_{(38,44)} = \mu+ \err_{(38,44),\{1\}}+ \err_{(38,44),\{
2\}
}+ \err_{(38,44),\{1,2\}}\quad \mbox{and}\nonumber\\[-8pt]\\[-8pt]
X_{\bsip}&=&X_{(38,19)} = \mu+ \err_{(38,19),\{1\}}+ \err
_{(38,19),\{
2\}}+ \err_{(38,19),\{1,2\}}.\nonumber
\end{eqnarray}
Because $\bsi$ and $\bsip$ share a value for $i_1$, they have the same
random effect for the set $u=\{1\}$. That is,
$\err_{\bsi,\{1\}}=\err_{\bsip,\{1\}}$. This is the only effect that
they share and so $\cov(X_{\bsi},X_{\bsip}) = \sigma^2_{\{1\}}$. More
generally, suppose that two indices $\bsi$ and $\bsip$ satisfy
$i_j=i'_j$ for and only for $j\in u$. Then $X_{\bsi}$ and $X_{\bsip}$
share random effects $\err_{\bsi ,v}=\err _{\bsip,v}$ for all nonempty
$v\subseteq u$ and so $\cov(X_{\bsi},X_{\bsip}) = \sum
_{v\dvtx\varnothing\ne v\subseteq u}\sigma_v^2$.\vspace*{1pt}

The expression $\err_{(38,44),\{1\}}$ is mildly redundant
since the second index $i_2=44$ is ignored.
We could have written it as $\err_{(38),\{1\}}$.
Such a choice amounts to writing the
general case as $\err_{\bsi_u,u}$, which is
more cumbersome when it appears in lengthy expressions.

The sample mean of $X$ is the ratio
%
%
\begin{equation}\label{eqxbar}
\bar X=
\sbsi X_{\bsi}Z_{\bsi}\Big/ \sbsi Z_{\bsi},
\end{equation}
where the sums are over all
index values $\bsi$.
The denominator in (\ref{eqxbar}) is
the total number $N$ of observations.
Our goal is to estimate the variance
of $\bar X$ by resampling methods.

\subsection{Partial duplicate observations}

We will need to keep track of the
extent to which different observations
have the same index values, in order
to properly reflect correlations among
the $\xbsi$.

For each $\bsi$ and $u\subseteq[r]$,
the number
\[
\nbsiu= \sum_{\bsip}Z_{\bsip}\oiuipu\vadjust{\goodbreak}
\]
counts how many observations match $\xbsi$
for all indices $j\in u$.
If $\zbsi=1$, then $\nbsiu\ge1$
because $\xbsi$ matches itself.
By convention, $N_{\bsi,\varnothing} = N$
and $N_{\bsi,[r]}=1$.
The quantity
\[
\nu_{u} = \frac1N\sbsi\zbsi\nbsiu\ge1
\]
is the average number of matches in the
subset $u$ for observations in the data set,
and $\nu_{[r]}=1$.

The most important of the $\nu_u$ are
for singletons $u=\{j\}$. The value $\nu_{\{j\}}$
has a quadratic dependence on the pattern
of duplication in the data. To see this,
write $n_{\ell j}=\sbsi\zbsi1_{i_j=\ell}$
for the number of times that variable
$j$ is equal to $\ell$ in the data.
Then $\nu_{\{j\}} = N^{-1}\sum_{\ell=1}^\infty n_{\ell,j}^2$
because each $N_{\bsi,\{j\}} = n_{i_j,j}$
appears $n_{i_j,j}$ times in the summation defining
$\nu_{\{j\}}$.

If $u\subseteq v$, then $\nu_u \ge\nu_v$.
In some applications $\nu_u\gg\nu_v$
for proper subsets $u\subsetneq v$.
For those applications, multiple matches are very unusual.
In other settings two factors, say, $i_1$
and $i_2$, might be highly though not perfectly dependent
(e.g., customer ID and phone number)
and then $\nu_{\{1,2\}}$ might be only
slightly smaller than $\nu_{\{1\}}$
or $\nu_{\{2\}}$. We return to this issue
in Section~\ref{secinterp}.

The specific pair of data values
$\bsi$ and $\bsip$ match in components
\[
\miip= \{ j\in[r]\mid i_j=i'_j\}.
\]
For the motivating data,
most of the $\miip$ are empty
and most of the rest have cardinality $|\miip|=1$.
We have $|\miip|=r$
if and only if $\bsi=\bsip$.
Although~$\miip$ is defined for all pairs
$\bsi$ and $\bsip$, we only use it when
$\zbsi\zbsip=1$, that is, when both $\xbsi$
and $\xbsip$ have been observed,
and the term ``most'' above refers to these pairs.

For each $\bsi$ and $k=0,1,\ldots,r$, the number
\[
\nbsik= \sum_{\bsip}Z_{\bsip}\one_{|\miip|=k}
\]
counts how many observations match $\xbsi$
in exactly $k$ places.

\subsection{\texorpdfstring{Random effects variance of $\bar X$}{Random effects variance of X}}

Here we record the true variance of~$\bar X$,
using the random effects model.
This is the quantity we hope to estimate by
bootstrapping.
%
%
\begin{theorem}\label{thmreffvar}
In the random effects model (\ref{eqreff})
%
%
\begin{equation}
\var( \bar X ) = \frac1N\sneu\nu_u\sigma^2_u.
\end{equation}
\end{theorem}

The contributions of the variance
components $\sigma^2_u$ are
proportional to the duplication indices $\nu_u$.
For large sparse data sets
we often find that $1\ll\nu_u\ll N$ when $0<|u|<r$.\vadjust{\goodbreak}

Our bootstrap approximations to this
variance are centered around a~quantity
$(1/N)\sum_{u\ne\varnothing}\gamma_u\sigma^2_u$
for gain coefficients $\gamma_u$ that
depend on the data configuration and
the particular bootstrap method.
Ideally, we want $\gamma_u = \nu_u$.
More realistically, some bootstrap
methods are able to
get $\gamma_u\ge\nu_u$ with $\gamma_u$
just barely larger than $\nu_u$
for the singletons $u=\{j\}$
which we expect to dominate $\var(\bar X)$.

\section{Naive bootstrap methods}\label{secnaive}

There are two main ways to bootstrap: resampling
[\citet{efro1979}]
and reweighting [\citet{rubi1981}],
with the distinction being
that the former uses a multinomial
distribution on the data while the
latter applies independent random weights
to the observations.

Naive bootstrap methods simply
resample or reweight the $N$
observations without regard to their
factorial structure.
That is, they use the same bootstrap one might use
for IID samples.
Here we show that naive bootstrap resampling
and reweighting have very similar and
very unsatisfactory performance.

\subsection{Naive resampling}

In the naive bootstrap, all $N$
observations are resampled with replacement.
The naive bootstrap variance of $\bar X$ converges to
%
%
\begin{equation}\label{eqnbootvar}
\var_{\nb}(\bar X)
= \frac1{N^2} \sbsi\zbsi(\xbsi-\bar X)^2
\end{equation}
as the number of resampled data sets tends to infinity.
%
%
\begin{theorem}\label{thmerevnbxb}
Under the random effects model (\ref{eqreff}),
the expected value of the naive bootstrap
variance of $\bar X$ is
%
%
\begin{equation}\label{eqerevnbxb}
\e_{\re}(\var_\nb( \bar X))
=\frac1{N} \sneu\sigma^2_u\biggl(1 -\frac{\nu_u}{N}\biggr).
\end{equation}
\end{theorem}

When $r>1$, the naive bootstrap can severely underestimate the
coefficients of~$\sigsqsetj$. We can see the effect in the Netflix
data, which has $r=2$. The gain coefficients $\nu_u$ can be computed
directly. For a random variable $X$ following the random effects model
(\ref{eqreff}) with variance components $\sigma^2_{\mathrm{movies}}$,
$\sigma^2_{\mathrm{raters}}$ and
$\sigma^2_{\mathrm{movies}\times\mathrm{raters}}$ we have
\[
\var(\bar X) \doteq \frac1N
(56\mbox{,}200 \sigma^2_{\mathrm{movies}}+646 \sigma^2_{\mathrm
{raters}}+\sigma^2_
{\mathrm{movies}\times\mathrm{raters}}),
\]
while
\[
\var_{\nb}(\bar X) \le \frac1N
(\sigma^2_{\mathrm{movies}}+\sigma^2_{\mathrm{raters}}+\sigma
^2_{\mathrm
{movies}\times\mathrm{raters}}),
\]
where $N\doteq100\mbox{,}000\mbox{,}000$. If $X$ has large
variance components for movies, the underestimation
can be severe. Even quantities dominated by a rater
effect will have a naive bootstrap variance far too
small.\vadjust{\goodbreak}

Theorem~\ref{thmerevnbxb}
generalizes Lemma 2 of \citet{pbs} which
treats naive bootstrap sampling for $r=2$.
We note that Owen [(\citeyear{pbs}), page 391] has an
error: it gives
the coefficient of $\sigma^2_{\{1,2\}}$
as $1/N$ where it should be $1/(N-1)$.

\subsection{Naive reweighting}

Posterior sampling under the
Bayesian bootstrap [\citet{rubi1981}]
uses independent $\dexp(1)$ weights
on the sample values.
This corresponds to a posterior distribution
on $X$ that is a Dirichlet distribution with parameter
vector $(1,1,\ldots,1)$ with a $1$ for each
observation. The corresponding prior is a degenerate
Dirichlet with a parameter of $0$ on all possible
values for the random variable. The posterior
is degenerate, putting $0$ probability on any value
of $X$ that was never seen in the sample, thus
eliminating the user's need to know which possible
values were not in fact observed.
This motivation is simplest when the observations are
assumed to be distinct as, for example, with continuously
distributed values, but the method is also used on
data with ties.

In the naive Bayesian bootstrap, all
$N$ observations are given random weights which
are then normalized.
Observation $\bsi$ gets weight $W_{\bsi}\sim G$
independently sampled.
We assume that $G$ has mean $1$ and variance $\tau^2<\infty$.
Typically, $\tau^2=1$.

The original Bayesian bootstrap [\citet{rubi1981}]
had $W_{\bsi}\sim\dexp(1)$, but other
distributions are useful too.
Taking $W_{\bsi}\sim\dpoi(1)$ gives
a result very similar to the usual
bootstrap, and it has integer weights.
Independent $\dbin(N,1/N)$ weights would
provide a more exact match, but for large
$N$ there is no practical difference
between $\dbin(N,1/N)$ and $\dpoi(1)$.
See \citet{oza2001} and \citet{leeclyd2004}
for uses of independent reweighting in bagging and boosting.

Taking $W_{\bsi}\sim\dustd\{0,2\}
=(\delta_0+\delta_2)/2$
also has integer values. The algorithm
goes ``double or nothing'' independently
on all $N$ observations.
The nonzero integer values are all
equal, so these weights correspond
to using a random unweighted subset of the data.
Double-or-nothing weighting is then
a version of half-sampling methods [\citet{mcca1969}]
without the constraint on the sum of weights,
just as Poisson weighting removes a sum
constraint from the original bootstrap.

The choice of weights makes a small
difference to the bootstrap performance.
See Section~\ref{secbootstab}.

Each bootstrap resampled mean takes the form
\[
\bar X^* = T^*/N^*,
\]
where\vspace*{1pt}
$T^* = \sbsi\wbsi\zbsi\xbsi$
and $N^* = \sbsi\wbsi\zbsi$.
The bootstrap mean $T^*/N^*$ is a ratio
estimator of $\bar X$.
The asymptotic formula for the variance is
\[
\nbbvar(\bar X^*) = \frac1{N^2}
\e_{\nbb}\bigl(( T^*-\bar X N^*)^2\bigr).
\]
The tilde on $\var_{\nbb}$ is a reminder that this
formula is a delta method approximation: it is
the variance of a Taylor approximation to
$\bar X^*$. Because~$N$ is usually very large in the target applications,
we consider $\wt\var_{\nbb}$ to be a~reliable
proxy for $\var_{\nbb}$.
%
%
\begin{theorem}\label{thmnwvar}
In the random effects model (\ref{eqreff})
%
%
\begin{equation}\label{eqnwvar}
\e_{\re}( \nbbvar(\bar X^*))
= \frac{\tau^2}N\sneu\sigma_u^2\biggl( 1 - \frac{\nu_u}N\biggr).
\end{equation}
\end{theorem}

The naive Bayesian bootstrap using $\tau^2=1$
has the same average variance as the naive bootstrap.
In large data sets we may find that $\nu_u\gg\tau^2(1-\nu_u/N)$
and then the Bayesian bootstrap greatly underestimates
the true variance. When $\max_{u\ne\varnothing}\nu_u$
is not too large, then Theorem~\ref{thmnwvar}
offers a way to counter this problem. We can simply multiply the naive
bootstrap variance
by $\tau^2 = \max_{u\ne\varnothing}\nu_u$
to get conservative variance estimates.
The largest $\nu_u$ comes from $u=\{j\}$
for some $j\in[r]$ and it is an easy quantity
to compute.

\subsection{Bootstrap stability}\label{secbootstab}

Any distribution on weights
with $\e(W)=1$ and $\var(W)=\tau^2$
will have the same value for
$\e_{\re}(\wt\var_{\nbb}(\bar X^*))$.
Several different weight distributions
are popular in the literature.
\citet{oza2001}
takes $W_{ij}$ to be independent Poisson random
variables with mean $1$. This creates
a~very close approximation to the original bootstrap's
multinomial weights.
\citet{leeclyd2004}
prefer exponential weights with mean $1$
because they yield an exact online version of the
Bayesian bootstrap.

In this section we look at the effect of
the weight distribution. Other things being
equal, we prefer a bootstrap to yield a stable
variance estimate. That is, we like a smaller
variance under bootstrap sampling for the
estimated variance of the mean.
For this purpose it is better to have weights
with a~small kurtosis
$\kappa=\e((W-1)^4)/\tau^4-3$.
The smallest possible kurtosis for weights
with mean $1$ and variance~$1$ arises
for weights uniformly distributed on the
values $0$ and $2$.
The kurtosis of the data,
$\kappa_x=(1/N)\sbsi\zbsi(\xbsi-\bar X)^4/\sigma^4-3$
where $\sigma^2=(1/N)\sbsi\zbsi(\xbsi-\bar X)^2$,
also plays a role. We work
out the consequences for the naive bootstrap
for simplicity.

If we hold the observations $\xbsi$ fixed and
implement the bootstrap, doing some
number $B$ of replicates, we will
estimate the quantity
\[
\nbbvar(\bar X^*)
=\frac1{N^2}\sbsi\sbsip\zbsi\zbsip\e_{\nbb}(\wbsi\wbsip)\ybsi
\ybsip,
\]
where $\ybsi= \xbsi-\bar X$.
To estimate this variance, we may use
%
%
\begin{eqnarray}
\label{eqhathat}
\wh{\wt\var}_{\nbb}(\bar X^*)
&=&\frac1{BN^2}\sum_{b=1}^B\sbsi\sbsip\zbsi\zbsip\wbsib\wbsipb
\ybsi
\ybsip\nonumber\\[-8pt]\\[-8pt]
&=&\frac1{B}\sum_{b=1}^B\biggl(\frac1N\sbsi\zbsi\wbsib(\xbsi
-\bar
X)\biggr)^2,\nonumber
\end{eqnarray}
where $\wbsib$ are independent identically distributed
random weights and $b=1,\ldots,B$ indexes
the bootstrap replications.
The hat in (\ref{eqhathat}) represents
estimation from $B$ bootstrap samples.
It is possible to use (\ref{eqhathat}) with $B=1$.
That such a ``unistrap'' is possible stems
from the use of a delta method approximation.

Equation (\ref{eqhathat}) is not the usual
estimator. The more usual variance estimate is
%
%
\begin{equation}
s^2_{\nbb}(\bar X^*)=\frac1{B-1}\sum_{b=1}^B(\bar X^*_b-\bar
X^*_\sumdot)^2,
\end{equation}
where
%
%
\begin{equation}
\bar X^*_b = \frac1N\sbsi\zbsi\wbsib\xbsi
\quad\mbox{and}\quad
\bar X^*_\sumdot= \frac1B\sum_{b=1}^B\bar X^*_b.
\end{equation}
%
%
\begin{theorem}\label{thmstability}
Let $W$ and $\wbsib$ be IID random variables
with mean $1$ variance $\tau^2$ and kurtosis $\kappa_w<\infty$.
Then holding $\ybsi=\xbsi-\bar X$ fixed,
\[
\var_{\nbb}(\wh{\wt\var}_{\nbb}(\bar X^*)) = \frac{\sigma^4\tau
^4}{BN^2}
\biggl(2 + \frac{\kappa(\kappa_x+3)}{N}\biggr),
\]
where $\sigma^2 = (1/N)\sbsi\zbsi\ybsi^2$
and $\kappa_x = (1/N)\sbsi\zbsi\ybsi^4/\sigma^4-3$.
A delta method approximation gives
\[
\var_{\nbb}(s^2_{\nbb}) \doteq\frac{\sigma^4\tau^4}{BN^2}
\biggl(\frac{2B}{B-1} + \frac{\kappa(\kappa_x+3)}{N}\biggr).
\]
\end{theorem}

When $\kappa(\kappa_x+3)\ll N$, then
$\wh{\wt\var}_{\nbb}(\bar X^*)$
with $B$ reweightings has approximately the variance
of $s^2_{\nbb}(\bar X^*)$ with $B+1$ reweightings.

We find here that there are only small differences
between weighting schemes, but double-or-nothing
weights having the smallest possible kurtosis
$\kappa=-2$
have the best stability.
The $\dpoi(1)$ distribution has $\kappa=1$ and the
$\dexp(1)$ distribution has $\kappa=6$.

\section{Factorial reweighting}\label{secproposal}

Our proposal here is to apply a product of independent random weights
to the data. Observation $\bsi$ is given weight \mbox{$W_{\bsi}\ge0$}. The
weights take the form
%
%
\begin{equation}\label{eqprodwts}
W_{\bsi} = \prod_{j=1}^r W_{j,i_j},
\end{equation}
where $W_{j,i_j}$ are independent random variables for $j\in[r]$ and
$i_j\ge1$. We assume that $\e(W_{j,i_j})=1$ and
$\var(W_{j,i_j})=\tau^2_j<\infty$. The usual choice has all~$\tau_j^2$
equal to a common $\tau^2$ which in turn is usually equal to $1$.

For the example in equation (\ref{eqexample38}), the observation at
index $\bsi= (38,44)$ gets weight $\wbsi= W_{1,38}W_{2,44}$. It shares
one weight factor with the observation at $\bsip=(38,19)$ which has
$\wbsip= W_{1,38}W_{2,19}$.\vspace*{1pt}

The reweighted mean $\bar X^*$ is once again
a ratio estimate with delta method approximation
%
%
\begin{equation}\label{eqvarhat}
\wt\var_{\bpb}(\bar X^*) = \frac1{N^2}
\e_{\bpb}\bigl(( T^*-\bar X N^*)^2\bigr),
\end{equation}
where $T^*=\sbsi\zbsi\wbsi\xbsi$ and
$N^*=\sbsi\zbsi\wbsi$ for $\wbsi$ given
by (\ref{eqprodwts}).
The subscript $\bpb$ refers to random weights
taking the product form.

The bootstrap variance
depends on precise details of the overlaps among
different observations.
We will derive some approximations to this
variance below.
For the exact variance
we need to introduce some additional quantities:
\begin{eqnarray*}
\rho_k & = &\frac1{N^2}\sbsi\sbsip\zbsi\zbsip\one_{|\miip|=k},\\
\nu_{k,u} & = &\frac1{N}\sbsi\sbsip\zbsi\zbsip
\one_{|\miip|=k}\oiuipu
\end{eqnarray*}
and
\begin{eqnarray*}
\wt\nu_{k,u}
& = &\frac1{N^2}\sbsi\sbsip\sbsipp\zbsi\zbsip\zbsipp
\one_{|\miip|=k}\one_{\bsi_u=\bsippp_u}\\
& = &\frac1{N^2}\sbsi\zbsi\nbsiu\nbsik
\end{eqnarray*}
for $k=0,1,\ldots,r$ and $u\subseteq[r]$.
In words, $\rho_k$ gives the fraction of data pairs
that match in exactly $k$ positions,
while $\nu_{k,u}/N$ gives the fraction of data pairs
that match in exactly $k$ positions including
all $j\in u$. The third quantity,~$\wt\nu_{k,u}$,
is $N$ times the fraction of data triples $(\bsi,\bsip,\bsipp)$
in which $\bsi$ matches $\bsip$ in precisely
$k$ places while also matching $\bsipp$ for all $j\in u$.

These new quantities satisfy the identities
\[
\sum_{k=0}^r\rho_k = 1 \quad\mbox{and}\quad
\sum_{k=0}^r\nu_{k,u}=
\sum_{k=0}^r\wt\nu_{k,u}=\nu_u.
\]
Also, it is clear that $\nu_{k,u}=0$ when $|u| > k$.
%
%
\begin{theorem}\label{thmpwvar}
In the random effects model (\ref{eqreff})
%
%
\begin{equation}\label{eqpwvar}
\e_{\re}( \wt\var_{\bpb}(\bar X^*))
= \frac1N\sneu\gamma_u\sigma^2_u,\vadjust{\goodbreak}%
\end{equation}
where
%
%
\begin{equation}\label{eqdefgain}
\gamma_u=\sum_{k=0}^r(1+\tau^2)^k
(\nu_{k,u}-2\wt\nu_{k,u}+\rho_k\nu_u).
\end{equation}
\end{theorem}

The quantities $\gamma_u$ are ``gain coefficients''
which multiply $\sigma^2_u/N$. Ideally they should
equal $\nu_u$ and then the bootstrap variance
would match the desired one. Where they differ
from $\nu_u$, the bootstrap variance is biased.
Typically, the bias is positive, making this bootstrap
conservative. Sometimes the bias is very small.

The special\vspace*{1pt} case $r=1$ is interesting
because it corresponds to IID sampling.
Then the only variance component
is $\sigma^2_{\{1\}}$, which we
abbreviate to $\sigma^2$
and equation (\ref{eqpwvar})
simplifies to
\[
\frac{\tau^2\sigma^2}N\biggl(1-\frac1N\biggr)
=\frac{\tau^2\sigma^2}{N-1}.
\]
In this instance there is a (trivial) negative
bias if $\tau^2=1$.

Independently reweighting rows and columns is similar to independently
resampling them. That strategy of bootstrapping
rows and columns has been given several names in the literature.
\citet{brenharrhans1987} called it ``boot-p,i'' because for
educational testing data, it resamples both people and items. McCullagh
[(\citeyear{mccu2000}), page 294] calls the method ``\mbox{Boot-II}.'' There
is also another ``Boot-II'' for the one way layout in that paper.
Noting a similarity to Cornfield and Tukey's pigeonhole model for
analysis of variance, \citet{pbs} calls this approach the
``pigeonhole bootstrap.'' Reweighting with a product of Rubin's
(\citeyear{rubi1981}) exponential weights is thus a~``Bayesian
pigeonhole bootstrap.''


\section{Interpretable approximations}\label{secinterp}

Theorem~\ref{thmpwvar} gives
exact finite sample formulas for the gain
coefficients $\gamma_u$,
but they are unwieldy. Here we make some
approximations to $\gamma_u$
in order to get more interpretable results.

First we introduce the quantity
\[
\eps
= \max_{\bsi}\max_{u\ne\varnothing}\frac{\nbsiu}N
= \max_{\bsi}\max_{1\le j\le r}\frac{N_{\bsi,{\{j\}}}}N,
\]
which measures the largest proportional duplication
of indices.
Though $1\ge\eps\ge1/N$,
we anticipate that $\eps$ will usually be small.
For the Netflix data,
$\eps= 232\mbox{,}944/100\mbox{,}480\mbox{,}507\doteq0.00232$,
stemming from one movie having $232\mbox{,}944$ ratings.

Although we suppose that $\eps$ is small
below, it is worth pointing out that
exceptions do arise, even for some very
large data sets. For example, if
the observed data form a complete
$N_1\times N_2\times\cdots\times N_r$
sample, then $\eps= \max_{1\le j\le r}1/N_j$.
If one factor takes only a modest number of
levels, then $\eps$ is large.
A second context where $\eps$ is large arises
when one of the factors is greatly dominated
by one of its levels, as, for example, we
might find in Internet data where one factor
is the country of the web user.

A second parameter to aid interpretability is
\[
\eta= \max_{\varnothing\subsetneq u\subsetneq v}\frac{\nu_v}{\nu_u}.
\]
By construction $\eta\le1$, and
we ordinarily expect $\eta$ to be small.
Of the indices which match for $j\in u$,
only a relatively small number should also match for $j \in v-u$
too, because each additional match in large
data sets represents a~coincidence.
For the Netflix data
\[
\eta= \max\bigl\{
\nu_{\{1,2\}}/\nu_{\{1\}},
\nu_{\{1,2\}}/\nu_{\{2\}}\bigr\}=1/646\doteq0.00155.
\]

While $\eta$ is often small,
there are exceptions. If
two factors are very dependent, then $\eta$
need not be small. For example, people's names
and phone numbers may be such variables:
many or even most phone numbers are used by a small number
of people (often one) and many people use only a small
number of phone numbers. Then the fraction of
data pairs matching on both of these variables will
not be much smaller than the fraction matching on one of them.

In simplifying expressions we use
$O(\eta)$ and $O(\eps)$. These describe
limits as~$\eta$ (resp., $\eps$)
converge to $0$. The implied constants
may depend on $r$.
In some expressions we have retained
explicit constants.
%
%
\begin{theorem}\label{thmbpbgains}
In the random effects model (\ref{eqreff}),
the gain coefficient (\ref{eqdefgain}) for $u\ne\varnothing$
in the product reweighted bootstrap is
%
%
\begin{equation}\label{eqbpbgains}
\gamma_u = \nu_u\bigl[ (1+\tau^2)^{|u|}-1
+\theta_u\err\bigr] + \sum_{v\supsetneq u}(1+\tau^2)^{|v|}(\tau
^2)^{|v-u|}\nu_v,
\end{equation}
where $|\theta_u|\le(1+\tau^2)((1+\tau^2)^r-1)/\tau^2$.
For $\tau^2=1$,
\[
\gamma_u = \nu_u\bigl[ 2^{|u|}-1+\theta_u\err\bigr]
+\sum_{v\supsetneq u}2^{|v|}\nu_v,
\]
where $|\theta_u|\le2^{r+1}-2$.
\end{theorem}

For $r=2$ using $\nuset{\{1,2\}}=1$
and the usual choice $\tau^2=1$,
we find that
\[
\gamsetj
=\nusetj\bigl(1+\theta_{\{j\}}\eps\bigr)+2,\qquad
j=1,2,
\]
and
\[
\gamset{\{1,2\}}
=\nuset{\{1,2\}}\bigl(3+\theta_{\{1,2\}}\eps\bigr),
\]
where each $|\theta|\le6$.
The Bayesian pigeonhole bootstrap variance
closely matches the ordinary
pigeonhole bootstrap variance.
In the extreme\vadjust{\goodbreak} setting where
$\sigma^2_{\{1\}}=\sigma^2_{\{2\}}=0<\sigma^2_{\{1,2\}}$
the resulting\vspace*{1pt} bootstrap variance is about
three times as high as it should be.
In a limit as $\min_j\nu_{\{j\}}\to\infty$
and $\eps\to0$,
%
%
\begin{equation}\label{eqitworks}
\frac{\e_{\re}( \wt\var_{\bpb}(\bar X^*))}{\var
(\bar X)}\to1
\end{equation}
holds for fixed $\sigma^2_{\{j\}}>0$, $j=1,2$.
For $r=3$, with $\nuset{\{1,2,3\}}=1$ and $\tau^2=1$,
\begin{eqnarray*}
\gamset{\{1\}}
&\approx&
\nuset{\{1\}}+4\nuset{\{1,2\}}+4\nuset{\{1,3\}}+8,\\
\gamset{\{1,2\}}
&\approx& 3\nuset{\{1,2\}}+8\quad
\mbox{and}\quad
\gamset{\{1,2,3\}} \approx 7,
\end{eqnarray*}
where $\approx$ reflects an additive error
of size $\nu_u\theta_u\err$ for $|\theta_u|\le14$.
In the extreme case where the only nonzero
variance coefficient is $\sigma^2_{[3]}$, then
the product reweighted bootstrap variance is
about $7$ times as large as it should be.
On the other hand, when the main effect
variances $\sigma^2_{\{j\}}$ are positive
and $\nu_v/\nu_u\to0$ for $v\subsetneq u$,
then (\ref{eqitworks}) holds.
More generally, we have Theorem~\ref{thmintbounds}.
%
%
\begin{theorem}\label{thmintbounds}
For the random effects model (\ref{eqreff})
and the product reweight\-ed bootstrap
with $\tau^2=1$, the gain coefficient for nonempty $u\subseteq[r]$
satisfies
\[
2^{|u|}-1-(2^{r+1}-2)\eps<
\frac{\gamma_u}{\nu_u}\le
2^{|u|}(1+2\eta)^{|v-u|}-1 +(2^{r+1}-2)\eps.
\]
If there exist $m$ and $M$ with $0<m\le\sigma^2_u\le M<\infty$
for all $u\ne\varnothing$, then
\[
\frac{\e_{\re}( \wt\var_{\bpb}(\bar X^*))}{\var
(\bar X)}
= 1+O(\eta+\eps).
\]
\end{theorem}

The first claim of Theorem~\ref{thmintbounds}
can be summarized as
\[
\frac{\gamma_u}{\nu_u} = \bigl(2^{|u|}-1\bigr)\bigl(1+O(\eta)\bigr)+O(\eps)
\approx2^{|u|}-1,
\]
and the second as
${\e_{\re}( \wt\var_{\bpb}(\bar X^*))}/{\var(\bar X)}
\approx1$,
where the implied constants depend on $r$.
They generally grow exponentially in $r$
but the interesting values of $r$ are small
integers from $2$ to $6$ or so.
The main effects dominate when $\eta$
is small and they are properly accounted
for when $\eps$ is small.

\section{The heteroscedastic model}\label{sechet}

In the $r$-fold crossed random effects mod\-el~(\ref{eqreff}),
the term $\err_{\bsi,u}$ has the same variance for all $\bsi$.
This model may not be realistic. For instance, the Netflix
data includes some customers whose ratings have very small
variance and others with a very large variance. Similarly,
but to a lesser extent, movies also differ in the variance
of their ratings.
Unequal variances have the potential to bias
inferences, especially in unbalanced cases,
because the entities with more observations on
them might have systematically higher variance
than the others do.

A more realistic model is the
\textit{heteroscedastic $r$-fold crossed random
effects model}, with
%
%
\begin{equation}\label{eqreffhet}
X_{\bsi}
=\mu+ \sum_{u\ne\varnothing} \err_{\bsi,u},
\end{equation}
where $\mu\in\real$ and $\err_{\bsi,u}$
are independent random variables with
mean $0$ and variance~$\sigma^2_{\bsi,u}$.
There are more variance parameters than observations,
we do not need to estimate them.
\citet{pbs} gives conditions under which
the pigeonhole bootstrap with $r=2$
produces a variance estimate with relative
error tending to zero in the heteroscedastic setting.
Here we investigate product
reweighting with general $r$ for model (\ref{eqreffhet}).

We need some new quantities.
For $u\ne\varnothing$, define
\begin{eqnarray*}
\nubsiu& = &\frac1N\sbsip\zbsip\one_{\bsi_u=\bsip_u}=\frac
{\nbsiu}N,\\
\nu_{\bsi,k} &=& \frac1N\sbsip\zbsip\one_{|\miip|=k}=\frac
{\nbsik}N
\end{eqnarray*}
and
\[
\nu_{\bsi,k,u} = \frac1N\sbsip\zbsip\one_{|\miip|=k}\one_{\bsi
_u=\bsip_u}.
\]
We also will use
\[
\nusigbar = \frac1N\sbsi\zbsi\nubsiu\sigma^2_{\bsi,u}
\]
and
\[
\nukbar = \frac1N\sbsi\zbsi\nu_{\bsi,k}.
\]

Next, we parallel the development from
the ordinary random effects mod\-el~(\ref{eqreff}).
Theorem~\ref{thmreffvarhet}
gives the exact variance of $\bar X$
for heteroscedastic random effects,
Theorem~\ref{thmpwvarhet}
gives the gain coefficients
under product reweighting,
Theorem~\ref{thmbpbgainshet}
provides interpretable bounds for the
gains in terms of $\eps$. Finally,
Theorem~\ref{thmintboundshet} gives
conditions under which the product
reweighted bootstrap has a negligible
bias.
%
%
\begin{theorem}\label{thmreffvarhet}
In the heteroscedastic random effect model (\ref{eqreffhet})
%
%
\begin{equation}
\var( \bar X ) = \frac1N\sneu\sbsi\nubsiu\sigma^2_{\bsi,u}.
\end{equation}
\end{theorem}
%
%
\begin{theorem}\label{thmpwvarhet}
In the heteroscedastic random effects model (\ref{eqreffhet})
%
%
\begin{equation}\label{eqpwvarhet}
\e_{\re}( \wt\var_{\bpb}(\bar X^*))
= \frac1N\sneu\sbsi\gamma_{\bsi,u}\sigma^2_{\bsi,u},\vadjust{\goodbreak}%
\end{equation}
where
%
%
\begin{equation}\label{eqdefgainhet}
\gamma_{\bsi,u}
=\sum_{k=0}^r(1+\tau^2)^k
(\nu_{\bsi,k,u} -2\nu_{\bsi,k}\nu_{\bsi,u}+\nukbar\nu_{\bsi,u}).
\end{equation}
\end{theorem}
%
%
%
\begin{theorem}\label{thmbpbgainshet}
In the heteroscedastic random effects model (\ref{eqreffhet}),
the gain coefficient $\gamma_{\bsi,u}$
of (\ref{eqdefgainhet})
for $\zbsi=1$ and $u\ne\varnothing$
in the product reweighted bootstrap is
\[
\gamma_{\bsi,u} = \nu_{\bsi,u}\bigl[ (1+\tau^2)^{|u|}-1
+\theta_u\err\bigr] + \sum_{v\supsetneq u}(1+\tau^2)^{|v|}(\tau
^2)^{|v-u|}\nu
_{\bsi,v},
\]
where
$|\theta_u|\le(1+\tau^2)((1+\tau^2)^r-1)/\tau^2$.
For $\tau^2=1$
\[
\gamma_{\bsi,u} = \nu_{\bsi,u}\bigl[ 2^{|u|}-1
+\theta_u\err\bigr] + \sum_{v\supsetneq u}2^{|v|}\nu_{\bsi,v},
\]
where
$|\theta_u|\le2^{r+1}-2$.
\end{theorem}

Theorem~\ref{thmbpbgainshet} establishes
that our bootstrap is conservative in
the heteroscedastic case. With $\tau^2=1$
we have
\[
\frac{\gamma_{\bsi,u}}
{\nu_{\bsi,u}} \ge2^{|u|}-1-(2^{r+1}-2)\eps.
\]

For the homoscedastic random effects model,
the main effects dominate when
$\eta= \max_{\varnothing\subsetneq u\subsetneq v}{\nu_v}/{\nu_u}$
is small and the variance components are all
within the interval $[m,M]$ for $0<m\le M>\infty$.
In the heteroscedastic case we might reasonably
require every $\sigma^2_{\bsi,u}\in[m,M]$.
The analysis we used for Theorem~\ref{thmintbounds}
also requires the quantities
\[
\eta_{\bsi}
= \cases{
\displaystyle \max_{\varnothing\subsetneq u\subsetneq v}\frac{\nu_{\bsi,v}}{\nu
_{\bsi
,u}}, &\quad$\zbsi
=1$,\vspace*{2pt}\cr
0, &\quad$\zbsi=0$,}
\]
to be small.

For $r=2$ the only subsets $u$ and $v$
which appear in $\eta_{\bsi}$ are $u=\{j\}$
and $v=\{1,2\}$. Furthermore,
$\nu_{\bsi,\{1,2\}}=1/N$ and so
\[
\max_{\bsi} \eta_{\bsi}
=
\max_{j\in\{1,2\}}\max_{\bsi}\frac{N_{\bsi,\{j\}}}N
=\eps.
\]
Then using the same argument we used to prove
the second part of Theorem~\ref{thmintbounds},
we get
\[
\frac{\e_{\re}( \wt\var_{\bpb}(\bar X^*))}{\var
(\bar X)}
= 1+O(\eps)\qquad \mbox{for $r=2$}.
\]

The case for $r>2$ is more complicated.
There may be observations $\bsi$
with large values for $\nu_{\bsi,v}/\nu_{\bsi,u}$
where $\varnothing\subsetneq u\subsetneq v$.
We still get a good approximation
from the product reweighted bootstrap
because even though the individual~$\eta_{\bsi}$
need not always be small, sums of $\nu_{\bsi,v}$
over $i$ are small compared
to corresponding sums of $\nu_{\bsi,u}$ for $\varnothing\subsetneq
u\subsetneq v$.
%
%
\begin{theorem}\label{thmintboundshet}
For the heteroscedastic random effects model (\ref{eqreffhet}),
assume that there exist $m$ and $M$ with
$0<m\le\sigma^2_{\bsi,u}\le M<\infty$.
Then the product reweighted bootstrap with $\tau^2=1$
satisfies
\[
\frac{\e_{\re}(
\wt\var_{\bpb}(\bar X^*))}{\var(\bar X)}
= 1+O(\eta+\eps).
\]
\end{theorem}

\section{Nested random effects}\label{secnesting}

The $r$-fold crossed random effects model (\ref{eqreff})
excludes replicated observations by definition:
there can be only one $\xbsi$ for any combination
$\bsi$ of factors.
If two $X$'s are observed to share all
index values~$i_j$, we can incorporate them by
introducing an $r+1$st index $i_{r+1}$ which
breaks the ties.
Conditionally on the effects of the
first $r$ indices, distinct replicates are independent.
That is,
$\sigma^2_u=0$ when $r+1\in u$
but $u\ne\{1,2,\ldots,r+1\}$.
The replicate index $i_{r+1}$ is a factor that is
nested within the first $r$ factors.

More generally, we could have $s$ additional
indices corresponding to factors
crossed with each other, but nested
within our $r$ outer factors.
Then the index $\bsi\in\{1,2,\ldots\}^{r+s}$
uniquely identifies a data point.
Ordinary replication has $s=1$.
The nesting structure means that
%
%
\begin{equation}\label{eqnested}
\sigma_u^2 = 0 \qquad\mbox{if } u\cap\{r+1,\ldots,r+s\}\ne
\varnothing
\mbox{ and }u\cap[r] \ne[r].
\end{equation}
In words, the effect $\eps_{\bsi,u}$ is $0$ if the
factors in $u$ include \textit{any} of the inner factors
without including \textit{all} of the outer factors.

When one factor is nested within another,
such as replicates within subjects, it is a
common practice to resample or reweight the
outer factor only. For example, the resampled
data set might contain resampled subjects
retaining the repeated measurements from
each of them.

In the nested setting,
the variance of $\bar X$ under
an $r+s$ factor version of
the random effects model (\ref{eqreff})
is still $(1/N)\sum_{u\ne\varnothing}\nu_u\sigma^2_u$,
although many of the $\sigma^2_u$ terms are zero.

For $\bsi\in[r+s]$ let
$\lfloor\bsi\rfloor= (i_1,\ldots,i_r)$
be the indices of its outer factors.
We can study these nested models by
introducing the variables
\[
T_{\lfloor\bsi\rfloor} = \sbsip\zbsip1_{\lfloor\bsip\rfloor
=\lfloor
\bsi\rfloor}\xbsip
\]
and
\[
M_{\lfloor\bsi\rfloor} = \sbsip\zbsip1_{\lfloor\bsip\rfloor
=\lfloor
\bsi\rfloor},
\]
so that the sample mean
\[
\bar X = \frac1N\sbsi\zbsi\xbsi=
\frac
{\sum_{\lfloor\bsi\rfloor}T_{\lfloor\bsi\rfloor}}
{\sum_{\lfloor\bsi\rfloor}M_{\lfloor\bsi\rfloor}}
\]
is an $r$-factor ratio estimator.

When the numbers $M_{\lfloor\bsi\rfloor}$
of replicates for each outer factor
vary,
we obtain a heteroscedastic
random effects model in the first $r$
variables.

\section{Example: Loquacity of Facebook comments}\label{secexample}

We present an analysis of national differences in comment length on
Facebook. In particular, Facebook users can share links with their
friends. Their friends, and the posting user, can comment on the
link. We compare the length of these comments produced by users in the
United States using the site in American English (\textit{US users}) and
those produced by users in the United Kingdom using the site in British
English (\textit{UK users}). 
We restrict the analysis to US and UK users commenting on links shared
by US and UK users. We additionally consider two
different modes by which users can comment: the standard web interface
to Facebook (\textit{web}) and an application for some touchscreen
mobile phones (\textit{mobile}).

We treat the logarithm of the number of characters in a comment as the
outcome in the following random effects model:
\[
X_{cm,\bsi} = \mu_{cm} + \sum_{u \neq\varnothing} \varepsilon
_{cm,\bsi,u},
\]
where $\mu_{cm}$ is the mean log characters for country $c$ in mode
$m$. Here the members of $\bsi$ are indexes for the user sharing the
link (\textit{sharer}), the user commenting on the link
(\textit{commenter}) and the canonicalized URL being shared (\textit{URL}).
By definition, no comments have $0$ characters, and so each $X$ in our
data set is well defined.

The data consist of $X_{cm,\bsi}$ for a sample of comments by US and UK
who are using Facebook in American and British English, respectively,
during a~short period in 2011. This sample includes 18,134,419 comments
by 8,078,531 commenters on 2,085,639 URLs shared by 3,904,715 sharers.
We examine whether these US and UK users post comments of different
lengths for both of the modes.
The duplication coefficients for this data are
\begin{eqnarray*}
\nu_{\mathrm{sh}} &\doteq& 17.71,\qquad \nu_{\mathrm{com}} \doteq
7.71,\qquad \nu_{\mathrm{url}}\doteq26\mbox{,}854.92,\\
\nu_{\mathrm{sh},\mathrm{com}} &\doteq& 5.92,\qquad \nu_{\mathrm
{sh},\mathrm{url}} \doteq12.91,\qquad
\nu_{\mathrm{com},\mathrm{url}} \doteq5.19
\end{eqnarray*}
and
\[
\nu_{\mathrm{sh},\mathrm{com},\mathrm{url}} \doteq 4.88.
\]
The coefficient for URLs is conspicuously
large, indicating that a naive bootstrap would
be very unreliable.\vadjust{\goodbreak}

The sample mean for a country and mode is
\[
\widehat{\mu}_{cm} = \frac{\sum_{\bsi} Z_{cm,\bsi} X_{cm,\bsi} }{
\sum_{\bsi
} Z_{cm,\bsi} }.
\]
We regard $\widehat{\mu}_{cm}$ as an estimate of $\mu_{cm}$ conditional on
the observed combinations of sharers, commenters and URLs.

The four sample means for both countries and both modes suggest that
the US users write longer comments than UK users when commenting on the
web ($\widehat{\mu}_{\mathrm{US},\mathrm{web}} = 3.62$,
$\widehat{\mu}_{\mathrm{UK},\mathrm{web}} = 3.55$), while UK users
write longer comments than US users when commenting via the selected
mobile interface ($\widehat{\mu}_{\mathrm{US},\mathrm{mobile}} = 3.5$,
$\widehat {\mu }_{\mathrm{UK}, \mathrm{mobile}} = 3.57$). Many
differences between US and UK users likely contribute to these observed
differences. Before searching for causes of these two differences, a
data analyst would likely want to quantify the evidence for the
existence and size of these differences. We test whether these two
pairs of means are likely to be observed given the null hypothesis of
no difference in comment length between the countries within each mode.

Using software for Hive [\citet{thusetal2009}], a Hadoop-based
map-reduce data warehousing and parallel computing environment, we can
compute each of these four means for a number of bootstrap reweightings
of the data, while visiting each observation only once. When visiting
an observation, the hashed identifiers for the factor levels for that
observation are each used as seeds to random number generators. This
allows all nodes to use the same $U\{0,2\}$ draw in computing the
product weight for all observations that share a particular factor level.
{Note that users can be both sharers and commenters. Since users can
comment on their own shared links, some observations could have the
same factor level identifier for both the sharer and commenter levels.
We use different portions of the hashed identifier so that the weights
for these two roles are not dependent.} For each reweighting, we
compute four reweighted sample means
\[
\widehat{\mu}_{cm}^* = \frac{\sum_{\bsi} Z_{cm,\bsi} W_{cm,\bsi}
X_{cm,\bsi
} }{ \sum_{\bsi} Z_{cm,\bsi} W_{cm,\bsi} },
\]
corresponding to $c\in\{\mathrm{US},\mathrm{UK}\}$ and
$m\in\{\mathrm{web},\mathrm{mobile}\}$.

\begin{figure}

\includegraphics{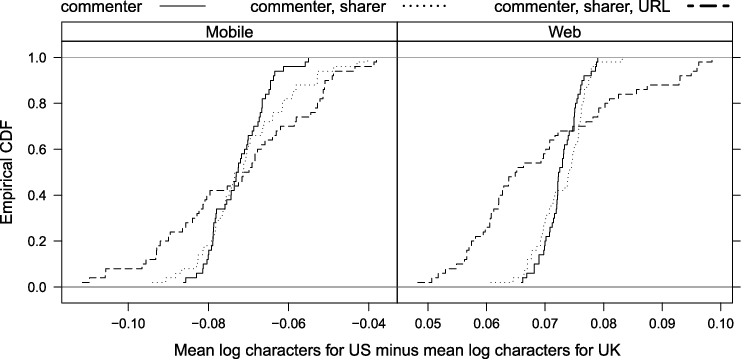}

\caption{Difference between the logged number of characters in comments
by US and UK users for three different bootstrap reweightings with
$R=50$. Each data point in the plotted ECDF is the difference in means
from a single bootstrap reweighting. US users post longer comments than
UK users on the web, but this difference is reversed for the mobile
interface studied.}
\label{factors-ecdf}
\vspace*{-2pt}
\end{figure}

\begin{figure}[b]

\includegraphics{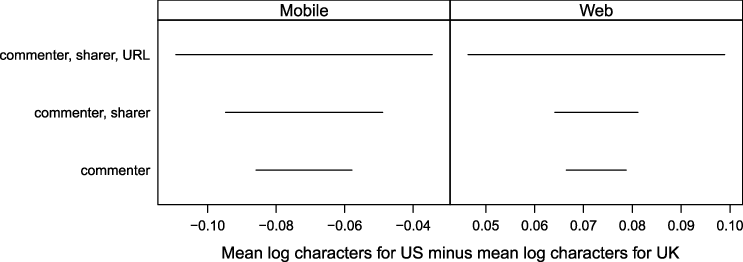}

\caption{Confidence intervals for the difference between the logged
number of characters in comments by US and UK users for three different
bootstrap reweightings with $R=50$. Confidence intervals span the 2.5\%
and 97.5\% quantiles of the normal with variance computed from the
bootstrap reweightings. While all three analyses reject the null
hypothesis, the one- and two-factor analyses may substantially
overstate confidence about the size of the true difference, especially
in the case of comments posted via the web interface.}
\label{factors-segplot}
\end{figure}

For comparison, we conduct this analysis reweighting one, two and all
three of the factors. Figure~\ref{factors-ecdf} presents $R=50$
bootstrapped differences in the two pairs of means when reweighting
commenters, commenters and sharers, and all three factors. Inspection
of these ECDFs confirms that the observed differences cannot be
attributed to chance, even when accounting for the random main and
interaction effects of commenters, sharers and URLs. The bootstrapped
differences in means are strikingly more dispersed for the three-factor
analysis. Figure~\ref{factors-segplot} shows 95\% confidence intervals
for the two differences computed as quantiles of the normal
distribution with variance computed from the bootstrap
reweightings.\vadjust{\goodbreak}
This highlights the substantial overstatement of certainty that can
come from neglecting the presence of additional random effects. In this
case, the three analyses would all reject the null hypothesis, but
would produce quite different confidence intervals.

%

%

For the approximations developed in Section~\ref{secinterp} to apply,
we require that~$\epsilon$ and~$\eta$ be small---that no single level
of any random effect make up a large portion of the observations and
that the number of observations matching on~$v$ is small compared to
the number matching on $u$ factors for all \mbox{$\varnothing\subsetneq u
\subsetneq v$}. We find that $\epsilon= 686\mbox{,}990 / 18\mbox
{,}134\mbox{,}419
\doteq0.0379$, as one URL had $686\mbox{,}990$ comments in this sample. We
also found that $\eta\doteq0.767$. Because $\eta$
is not very small it is possible that the variance estimates
are conservative.

\section{Discussion}\label{secdiscussion}

We have worked conditionally on
the observed values holding $\zbsi$ fixed.
It is clear that missingness can be informative
and thereby introduce a bias into a
sample mean.

The way to correct for missingness and even
whether to do such a correction is problem
dependent. In the Netflix data, the company
is seeking to predict ratings that were not
made and so the bias between observed and
unobserved ratings is of interest. The people
who competed in the Netflix contest were trying
to predict ratings that were actually made and
then artificially withheld, so the pairs
to be predicted were not subject to this bias.
For the Facebook data, some of the observed difference
between the lengths of comments by US and UK
users may be due to differences in which URLs they
comment on. An accounting of
missingness might involve inferring the likely
length of comments that would have been made
by US and UK users if they had the same propensity
to comment on particular URLs.
An analysis made conditionally on $\zbsi$
describes the statistical stability of comment lengths
for the actual pattern of commenting, which may be
of more interest.

To make an adjustment for missing data requires
some kind of assumption about the missingness
mechanism. That assumption cannot be tested
within a given data set because the necessary
confirmation values are not available. It is clear
that reweighting cannot correct a sampling
bias because many different sample biases may
be consistent with an observed data set.
In a~given problem with our preferred adjustment
for missingness built into the statistic
of interest, we could then consider how to
bootstrap the resulting bias adjusted statistics.
Alternatively, if the statistic is partially identified,
then we could consider how to bootstrap the resulting
sample bounds on the statistic.
It is not obvious which bootstrap method
would suit these tasks, but it seems clear that
in the random effects context product reweighting
will succeed more generally than naive bootstrapping.

We have used the variance of a sample
mean as a way to identify a suitable
bootstrap method. Plain sample
means are practically important. For example, click
through rates, or feature usage rates, are
means or ratios of means.
Even for this simple problem, naive bootstrapping
methods are severely downward biased in the random
effects setting. Product weighting replaces
this bias by a small upward bias that is more
acceptable in applications.

A bootstrap method
that underestimates the variance of
a mean cannot be expected to work well
on other problems. One that is properly
calibrated or conservative for the
variance of a scalar sample mean will
also work in some other settings.\vadjust{\goodbreak}

The extension to multivariate means
is very straightforward.
When $\xbsi\in\real^d$ for $d>1$ we may
replace the variances
$\sigma^2_u$ or $\sigma^2_{\bsi,u}$ by variance--covariance
matrices $\Sigma_u$ or $\Sigma_{\bsi,u}$, respectively,
in the variance formulas. This follows by
considering the variance of $\phi^\tran\xbsi$
for vectors $\phi\in\real^d$.

Bootstrap correctness extends from means
to other statistics. See
\citet{hall1992} and \citet{mamm1992}.
The extension
to smooth functions $g(\bar X)$ of means
is via Taylor expansion, when $g$
has a Jacobian matrix with full rank
at $\e(\bar X)$.

The bootstrap is usually used to get confidence intervals, not variance
estimates. For an asymptotically unbiased statistic that satisfies
a central limit \mbox{theorem}, a~properly calibrated variance yields
asymptotically correct bootstrap percentile confidence intervals. An
overestimated variance yields conservative percentile intervals.

Another way to extend from means to other
statistics is via estimating equations.
If the parameter $\widehat\theta$ is defined
by $\sbsi\zbsi m(\xbsi;\widehat\theta)=0$,
then we may test the hypothesis
that $\theta=\theta_0$ by testing
whether $m(\xbsi;\theta_0)$ has mean zero.
In practice, one would ordinarily form a histogram
of resampled $\widehat\theta^*$ values
and construct a confidence interval from them.

The heteroscedastic random effects model (\ref{eqreffhet})
has $2^r-1$ variance parameters for each observation.
Such a model can arise in an $r$-fold generalization
of factor analysis. Suppose that $F_{\bsi_u}$ is a
nonrandom factor depending on indices
in the set $u\subset\{1,2,\ldots,r\}$
and that $L_{\bsi_v}$ is a mean zero
random loading depending on indices
in the set $v\subset\{1,2,\ldots,r\}$ where
$u\cap v=\varnothing$.
Let
\[
X_{\bsi} = \mu+ \cdots+{F_{\bsi_u}L_{\bsi_v}} + \cdots+ \err
_{\bsi,\{
1,\ldots,r\}},
\]
where the ellipses hide other factors of the type just described
for different subsets of the variables.
The term shown contributes {$F_{\bsi_u}^2\var(L_{\bsi_v})$}
to 
the variance component for subset $v$ on observation $\bsi$.
Even if the loadings have constant variance, unequal factor
values will make this variance component heteroscedastic.
The factors and loadings could both have a product form
so that they contribute
\[
\prod_{j\in u}F_{j,i_j} \times\prod_{j\in{v}} L_{j,i_j}
\]
to $X_{\bsi}$ generalizing the SVD, but a product form is not necessary.

A generalized factor model would be extremely hard to
estimate. However, the total variance from all those different variance
contributions is handled by product reweighting, with
a small upward bias in the bootstrap variance of a mean.
A~similar phenomenon is well known in the context
of the wild bootstrap
[\citet{mamm1993}]
for the linear model. There a different distribution
is posited for each of $n$ observations in a regression
and the bootstrap process provides reliable inferences
for the regression coefficients without having to
accurately estimate all $n$ distributions.

%
\begin{appendix}
\section*{Appendix: Proofs}\label{app}

This Appendix contains theorem proofs
and a few lemmas.
The theorems are restated to make it easier
to follow the steps.
Equation numbers that appear in
the theorem statements from the
article are preserved in this Appendix.

\subsection*{Proof of Theorem \protect\ref{thmreffvar}}

\setcounter{theorem}{0}
\begin{theorem}
In the random effects model (\ref{eqreff})
\[
\var( \bar X ) = \frac1N\sneu\nu_u\sigma^2_u.
\]
\end{theorem}
\begin{pf}
The numerator of $\bar X$ in (\ref{eqxbar}) is
$\sbsi Z_{\bsi}X_{\bsi}
= N\mu+\sbsi\sneu Z_{\bsi}\err_{\bsi_u}$.
Therefore, the variance of $\bar X$ under
the random effects model is
\begin{eqnarray*}
\var(\bar X)
&=& \frac1{N^{2}}\e
\biggl(\sbsi\sbsip\zbsi\zbsip
\sneu\sneup\err_{\bsi,u}\err_{\bsi',u'}\biggr)\\
&=& \frac1{N^{2}}
\sneu\sigma^2_u
\sbsi\sbsip\zbsi\zbsip\oiuipu\\
&=& \frac1{N^{2}}\sneu\sigma^2_u\sbsi\zbsi\nbsiu\\
&=& \frac1{N}\sneu\nu_u\sigma^2_u.
\end{eqnarray*}
\upqed
\end{pf}

\subsection*{Proofs of Theorems \protect\ref{thmerevnbxb},
\protect\ref{thmnwvar} and \protect\ref{thmstability}}

Here we prove the theorems
about naive bootstrap sampling.
Theorem~\ref{thmerevnbxb} is about
naive resampling and
Theorem~\ref{thmnwvar} handles naive reweighting.
Theorem~\ref{thmstability} is about bootstrap
stability.
\begin{theorem}
Under the random effects model (\ref{eqreff}),
the expected value of the naive bootstrap
variance of $\bar X$ is
{\setcounter{equation}{6}
\begin{equation}
\e_{\re}(\var_\nb( \bar X))
=\frac1{N} \sneu\sigma^2_u\biggl(1 -\frac{\nu_u}{N}\biggr).
\end{equation}}
\end{theorem}
\begin{pf}
A $U$-statistic decomposition of the
sample variance is 
\begin{eqnarray*}
\var_\nb( \bar X)
&=&\frac1{2N^3} \sbsi\sbsip\zbsi\zbsip(\xbsi-\xbsip)^2\\
&=&\frac1{2N^3} \sbsi\sbsip\zbsi\zbsip\biggl( \sneu\err_{\bsi
,u}-\err_{\bsip,u}\biggr)^2.
\end{eqnarray*}
Under the random effects model
\begin{eqnarray*}
\e_{\re}(\var_\nb( \bar X))
&=&\frac1{2N^3} \sbsi\sbsip\zbsi\zbsip\sneu2\sigma^2_u(1-\oiuipu
)\\
&=&\frac1{N} \sneu\sigma^2_u\biggl(1 -\frac{\nu_u}{N}\biggr).
\end{eqnarray*}
\upqed
\end{pf}

To prove Theorem~\ref{thmnwvar},
we begin with a lemma on the covariance
of pairs of observations under the random
effects model.
%
%
\begin{lemma}\label{lemereyy}
Let $\xbsi$ follow the random effects model (\ref{eqreff})
and let $\ybsi=\break\xbsi-\bar X$.
Then
%
%
\setcounter{equation}{22}
\begin{equation}\label{eqexx}
\e_{\re}( \xbsi\xbsip) = \mu^2+\sneu\sigma^2_u\oiuipu
\end{equation}
and
%
%
\begin{equation}\label{eqeyy}
\e_{\re}( \ybsi\ybsip) = \sneu\sigma^2_u
\biggl(\oiuipu
-\frac{\nbsiu}N-\frac{\nbsipu}N+\frac{\nu_u}{N}\biggr).
\end{equation}
\end{lemma}
\begin{pf}
Equation (\ref{eqexx}) follows directly
from the random effects model definition.
Expanding
$\ybsi\ybsip$ yields
\[
\xbsi\xbsip
-\frac1N\sbsipp\zbsipp\xbsi\xbsipp
-\frac1N\sbsipp\zbsipp\xbsip\xbsipp
+\frac1{N^2}\sbsipp\sbsippp\zbsipp\zbsippp\xbsipp\xbsippp.
\]
Because $\mu$ cancels from $\ybsi$
we may assume that $\mu=0$ while
proving (\ref{eqeyy}).
Now
\[
\e_{\re}\biggl(\frac1N\sbsip\zbsip\xbsi\xbsip\biggr)
=\frac1N\sneu\sigma_u^2\sbsip\zbsip\one_{\bsi_u=\bsip_u}
=\frac1N\sneu\sigma_u^2\nbsiu.
\]
Therefore,
\[
\e_{\re}(\ybsi\ybsip)
=\sneu\sigma^2_u
\biggl(\oiuipu-\frac{\nbsiu}N-\frac{\nbsipu}N
+\frac1{N^2}\sbsipp\zbsipp\nbsippu
\biggr),
\]
which reduces to (\ref{eqeyy}).
\end{pf}
\begin{theorem}
In the random effects model (\ref{eqreff})
\setcounter{equation}{7}
\begin{equation}
\e_{\re}( \nbbvar(\bar X^*))
= \frac{\tau^2}N\sneu\sigma_u^2\biggl( 1 - \frac{\nu_u}N\biggr).
\end{equation}
\end{theorem}
\begin{pf}
Let $Y_i = X_i-\bar X$ and $T^*_y=\sbsi\wbsi\zbsi\ybsi$.
Then
\begin{eqnarray*}
\e_{\re}(\nbbvar(\bar X^*))
&=&\frac1{N^2}\e_{\re}
\bigl(\e_{\nbb}\bigl(( T^*-\bar X N^*)^2\bigr)\bigr)\\
&=&
\frac1{N^2}\e_{\re}\biggl(\sbsi\sbsip\zbsi\zbsip\ybsi\ybsip\e
_{\nbb
}(\wbsi\wbsip)\biggr)\\
&=&
\frac1{N^2}\sbsi\sbsip\zbsi\zbsip\e_{\re}(\ybsi\ybsip)\e
_{\nbb}(\wbsi
\wbsip).
\end{eqnarray*}
Next,
$ \e_{\nbb}(\wbsi\wbsip) = 1+\tau^2\one_{\bsi=\bsip}$.
Therefore,
%
%
\setcounter{equation}{24}
\begin{equation}\label{eqerevnwterms}
\e_{\re}(\nbbvar(\bar X^*))
=
\frac1{N^2}\sbsi\sbsip\zbsi\zbsip\e_{\re}(\ybsi\ybsip)
+
\frac{\tau^2}{N^2}\sbsi\zbsi\e_{\re}(\ybsi^2).\hspace*{-35pt}
\end{equation}
The double sum
in (\ref{eqerevnwterms}) vanishes because $\sbsi\zbsi\ybsi=0$.
Then from Lemma~\ref{lemereyy}, the coefficient of $\sigma_u^2$
in (\ref{eqerevnwterms}) is
\[
\frac{\tau^2}{N^2}\sbsi\zbsi
\biggl( 1 -\frac{2\nbsiu}{N}+\frac{\nu_u}N\biggr)
=
\frac{\tau^2}{N^2}
(N-2\nu_u+\nu_u),
\]
establishing (\ref{eqnwvar}).
\end{pf}
\begin{theorem}
Let $W$ and $\wbsib$ be IID random variables
with mean $1$ variance $\tau^2$ and kurtosis $\kappa_w<\infty$.
Then holding $\ybsi=\xbsi-\bar X$ fixed,
\[
\var_{\nbb}(\wh{\wt\var}_{\nbb}(\bar X^*)) = \frac{\sigma^4\tau
^4}{BN^2}
\biggl(2 + \frac{\kappa(\kappa_x+3)}{N}\biggr),
\]
where $\sigma^2 = (1/N)\sbsi\zbsi\ybsi^2$
and $\kappa_x = (1/N)\sbsi\zbsi\ybsi^4/\sigma^4-3$.
A delta method approximation gives
\[
\var_{\nbb}(s^2_{\nbb}) \doteq\frac{\sigma^4\tau^4}{BN^2}
\biggl(\frac{2B}{B-1} + \frac{\kappa(\kappa_x+3)}{N}\biggr).
\]
\end{theorem}
\begin{pf}
First, the variance of
$\wh{\wt\var}_{\nbb}(\bar X^*)$ scales as $1/B$
so we can work with $B=1$ and divide the result by $B$.
For $B=1$, we drop the subscript $b$ from $W$'s.
We will use the identity $\sbsi\zbsi\wbsi\ybsi= \sbsi\zbsi(\wbsi
-1)\ybsi$.
If $B=1$, then $\var_{\nbb}(\wh{\wt\var}_{\nbb}(\bar X^*))$
equals
\begin{eqnarray*}
&&\e_{\nbb}\biggl(\biggl(\sbsi\zbsi\wbsi\ybsi\biggr)^4\biggr)
-\biggl(\frac{\sigma^2\tau^2}{N}\biggr)^2\\
&&\qquad=\frac1{N^4}\sbsi\zbsi\e\bigl((\wbsi-1)^4\bigr)\ybsi^4
+\frac3{N^4}\sbsi\sbsip\zbsi\zbsip\e\bigl((\wbsi-1)^2\bigr)^2\ybsi
^2\ybsip^2\\
&&\qquad\quad{} -\frac3{N^4}\sbsi\zbsi\e\bigl((\wbsi-1)^2\bigr)^2\ybsi^4
-\biggl(\frac{\sigma^2\tau^2}{N}\biggr)^2\\
&&\qquad=
\frac{\tau^4\sigma^4(\kappa+3)(\kappa_x+3)}{N^3}
+\frac{3\tau^4\sigma^4}{N^2}
-\frac{3\tau^4\sigma^4(\kappa_x+3)}{N^3}-\frac{\sigma^4\tau
^4}{N^2}\\
&&\qquad=
\frac{\tau^4\sigma^4}{N^2}
\biggl(2 + \frac{\kappa(\kappa_x+3)}{N}\biggr).
\end{eqnarray*}

For the second part
\[
\var_{\nbb}(s^2_{\nbb}) = \e_{\nbb}(s^2_{\nbb})^2
\biggl(\frac{2}{B-1} + \frac{\kappa^*}{B}
\biggr),
\]
where $\kappa^*$ is the kurtosis of
$\bar X^*=
\sbsi\zbsi\wbsi\ybsi
/\sbsi\zbsi\wbsi$.\vspace*{1pt}
The delta method approximation to $\e_{\nbb}(s^2_{\nbb})$
is $\tau^2\sigma^2/N$.
For the kurtosis, we make the Taylor approximation
\[
\bar X^*
\doteq\bar X + \sbsi\zbsi(\wbsi-1)\ybsi.
\]
The expected value of $\bar X^*-\bar X$ reuses
much of the above computation and yields
\[
\e_{\nbb}\bigl((\bar X^*-\bar X)^4\bigr)
\doteq
\frac{\tau^4\sigma^4}{N^2}
\biggl(3 + \frac{\kappa(\kappa_x+3)}{N}\biggr).
\]
Therefore, $\kappa^*=\kappa(\kappa_x+3)/N$ and so
\[
\var_{\nbb}(s^2_{\nbb}) =
\frac{\tau^4\sigma^4}{BN^2}
\biggl(\frac{2B}{B-1} + \frac{\kappa(\kappa_x+3)}{N}
\biggr).
\]
\upqed
\end{pf}

\subsection*{Proofs of Theorems
\protect\ref{thmpwvar}, \protect\ref{thmbpbgains} and
\protect\ref{thmintbounds}}

Theorem~\ref{thmpwvar}
gives an exact expression for the
gain coefficients of the Bayesian pigeonhole
bootstrap in the constant variance
crossed random effects model.
Theorem~\ref{thmbpbgains}
gives an interpretable approximation
to those gain coefficients.
Theorem~\ref{thmintbounds} shows factorial reweighting
gives nearly the correct variance when
$\eps$ and $\eta$ are both small.
\begin{theorem}
In the random effects model (\ref{eqreff})
\setcounter{equation}{13}
\begin{equation}
\e_{\re}( \wt\var_{\bpb}(\bar X^*))
= \frac1N\sneu\gamma_u\sigma^2_u,%
\end{equation}
where
\begin{equation}
\gamma_u=\sum_{k=0}^r(1+\tau^2)^k
(\nu_{k,u}-2\wt\nu_{k,u}+\rho_k\nu_u).
\end{equation}
\end{theorem}
\begin{pf}
We begin along the same lines as Theorem~\ref{thmnwvar}
and find that
\[
\e_{\re}(\wt\var_{\bpb}(\bar X^*))
=
\frac1{N^2}\sbsi\sbsip\zbsi\zbsip\e_{\re}(\ybsi\ybsip)\e
_{\bpb}(\wbsi
\wbsip).
\]
For the product weights used in this bootstrap,
\[
\e_{\bpb}(\wbsi\wbsip) = \prod_{j\dvtx i_j=i'_j}(1+\tau^2)
=(1+\tau^2)^{|\miip|}
\]
with $\e_{\bpb}(\wbsi\wbsip)=1$ if $\bsi$ and $\bsip$
are not equal in any components.\vspace*{1pt}

From Lemma~\ref{lemereyy},
the coefficient of $\sigma^2_u$ in $\e_{\re}(\wt\var_{\bpb}(\bar
X^*))$ is
\begin{eqnarray*}
&&\frac1{N^2}\sbsi\sbsip\zbsi\zbsip
\biggl(\oiuipu-\frac{\nbsiu}N-\frac{\nbsipu}N
+\frac{\nu_u}{N}\biggr)(1+\tau^2)^{|\miip|}\\
&&\qquad=\frac1{N^2}\sbsi\sbsip\zbsi\zbsip
\biggl(\oiuipu-\frac{2\nbsiu}N
+\frac{\nu_u}{N}\biggr)(1+\tau^2)^{|\miip|}\\
&&\qquad=\frac1{N^2}
\sum_{k=0}^r(1+\tau^2)^k
\sbsi\sbsip\one_{|\miip|=k}\zbsi\zbsip
\biggl(\oiuipu-\frac{2\nbsiu}N
+\frac{\nu_u}{N}\biggr)\\
&&\qquad=
\frac1N\sum_{k=0}^r(1+\tau^2)^k
(\nu_{k,u}-2\wt\nu_{k,u}+\rho_k\nu_u).
\end{eqnarray*}
\upqed
\end{pf}

Next we establish an interpretable
approximation to the Bayesian pigeonhole
bootstrap variance, using the
quantity $\eps= \max_{\bsi}\max_jN_{\bsi,\{j\}}/N$
which is small unless the data are
extremely imbalanced.
\begin{theorem}
In the random effects model (\ref{eqreff}),
the gain coefficient (\ref{eqdefgain}) for $u\ne\varnothing$
in the product reweighted bootstrap is
\begin{equation}
\gamma_u = \nu_u\bigl[ (1+\tau^2)^{|u|}-1
+\theta_u\err\bigr] + \sum_{v\supsetneq u}(1+\tau^2)^{|v|}(\tau
^2)^{|v-u|}\nu_v,
\end{equation}
where $|\theta_u|\le(1+\tau^2)((1+\tau^2)^r-1)/\tau^2$.
For $\tau^2=1$,
\[
\gamma_u = \nu_u\bigl[ 2^{|u|}-1+\theta_u\err\bigr]
+\sum_{v\supsetneq u}2^{|v|}\nu_v,
\]
where $|\theta_u|\le2^{r+1}-2$.
\end{theorem}
\begin{pf}
The second claim follows immediately from the first
which we now prove.
We will approximate
$\gamma_u = \sum_{k=0}^r(1+\tau^2)^k
(\nu_{k,u}-2\wt\nu_{k,u}+\rho_k\nu_u)$.
First,
\begin{eqnarray*}
\sum_{k=0}^r(1+\tau^2)^k\nu_{k,u}
&=&\frac1N
\sum_{k=0}^r(1+\tau^2)^k\sbsi\sbsip\zbsi\zbsip\one_{|\miip
|=k}\one_{\bsi
_u=\bsip_u}\\
&=&\frac1N
\sum_{w\supseteq u}(1+\tau^2)^{|w|}\sbsi\sbsip\zbsi\zbsip\one
_{M_{ii'}=w}\\
&=&\frac1N
\sum_{w\supseteq u}(1+\tau^2)^{|w|}\sbsi\sbsip\zbsi\zbsip
\sum_{v\supseteq w}(-1)^{|v-w|}\one_{\bsi_w=\bsip_w}\\
&=&\sum_{w\supseteq u}(1+\tau^2)^{|w|}\sum_{v\supseteq
w}(-1)^{|v-w|}\nu_v.
\end{eqnarray*}
Writing $w\in[u,v]$ for $u\subseteq w\subseteq v$,
\begin{eqnarray*}
&& \sum_{w\supseteq u}(1+\tau^2)^{|w|}\sum_{v\subseteq
w}(-1)^{|v-w|}\nu_v\\
&&\qquad=\sum_{v\supseteq u}\nu_v\sum_{w\in[u,v]}(1+\tau
^2)^{|w|}(-1)^{|v-w|}\\
&&\qquad=\sum_{v\supseteq u}\nu_v\sum_{\ell=0}^{|v-u|}
{\pmatrix{|v-u|\cr\ell}}(-1)^\ell(1+\tau^2)^{|v|-\ell}\\
&&\qquad=\sum_{v\supseteq u}\nu_v(1+\tau^2)^{|v|}(\tau^2)^{|v-u|}.
\end{eqnarray*}

For the other parts of $\gamma_u$, we use
quantities $\theta$ that satisfy
bounds $0\le\theta\le1$. There
are several such quantities, distinguished
by subscripts, and defined at their
first appearance.
First, we have the bounds
%
%
\setcounter{equation}{25}
\begin{equation}
\frac{N_{\bsi,0}}N = 1- r\theta_{\bsi,0}\eps\quad \mbox
{and}\quad
\frac{\nbsik}N = \theta_{\bsi,k}\eps,\qquad 1\le k\le r.
\end{equation}
Next, for $u\ne\varnothing$,
\[
\wt\nu_{0,u}
= \frac1{N^2}\sbsi\zbsi\nbsiu N_{\bsi,0}
= \frac1N\sbsi\zbsi\nbsiu(1-r\theta_{\bsi,0}\eps)
= \nu_u(1- r \theta_{0,u}\eps)
\]
and for $k=1,\ldots,r$,
\[
\wt\nu_{k,u}
= \frac1{N^2}\sbsi\zbsi\nbsiu\nbsik
= \frac1N\sbsi\zbsi\nbsiu\theta_{\bsi,k}\eps
= \nu_u\theta_{k,u}\eps.
\]

Turning to $\rho_k$,
\[
\rho_0 = \frac1{N^2}\sbsi\zbsi N_{\bsi,0} =
\frac1{N}\sbsi\zbsi(1-\eps r\theta_{\bsi,0})=1-\eps r\theta
_0
\]
and
\[
\rho_k =
\frac1{N^2}\sbsi\zbsi\nbsik=
\frac1{N}\sbsi\zbsi\theta_{\bsi,k}\eps
=\theta_k\eps,\qquad k=1,\ldots,r.
\]

Now $-2\wt\nu_{0,u}+\rho_0\nu_u= -\nu_u
+\nu_u(2\theta_{0,u} -\theta_{0})r\eps$ and
\[
\sum_{k=1}^r(1+\tau^2)^k(-2\wt\nu_{k,u}+\rho_k\nu_u)
=
\nu_u\sum_{k=1}^r(1+\tau^2)^k(\theta_k-2\theta_{k,u})\eps.
\]
Therefore,
\[
\gamma_u
= \nu_u\bigl( (1+\tau^2)^{|u|}-1+\theta_u\eps\bigr)
+\sum_{v\supsetneq u}\nu_v(1+\tau^2)(\tau^2)^{|v-u|},
\]
where
\[
\theta_u = \sum_{k=1}^r(1+\tau^2)^k(\theta_k-2\theta_{k,u}).
\]
The proof follows because
$-1\le\theta_k-2\theta_{k,u}\le1$
and $\sum_{k=1}^r(1+\tau^2)^k = (1+\tau^2)((1+\tau^2)^r-1)/\tau^2$.
\end{pf}
\begin{theorem}
For the random effects model (\ref{eqreff})
and the product reweight\-ed bootstrap
with $\tau^2=1$, the gain coefficient for nonempty $u\subseteq[r]$
satisfies
\[
2^{|u|}-1-(2^{r+1}-2)\eps<
\frac{\gamma_u}{\nu_u}\le
2^{|u|}(1+2\eta)^{|v-u|}-1 +(2^{r+1}-2)\eps.
\]
If there exist $m$ and $M$ with $0<m\le\sigma^2_u\le M<\infty$
for all $u\ne\varnothing$, then
\[
\frac{\e_{\re}( \wt\var_{\bpb}(\bar X^*))}{\var
(\bar X)}
= 1+O(\eta+\eps).
\]
\end{theorem}
\begin{pf}
From Theorem~\ref{thmbpbgains}
\begin{eqnarray*}
\frac{\gamma_u}{\nu_u}&\le&
-1+\sum_{v\supseteq u}2^{|v|}\eta^{|v-u|} +(2^{r+1}-2)\eps\\
& = &2^{|u|}(1+2\eta)^{|v-u|}-1 +(2^{r+1}-2)\eps,
\end{eqnarray*}
and then using $\nu_v > 0$,
\[
\frac{\gamma_u}{\nu_u}>
2^{|u|}-1-(2^{r+1}-2)\eps.
\]

For the second claim, small $\eta$ means
that the variance is dominated by contributions
$\sigma^2_{\{j\}}$ for which $\gamma_{\{j\}}\approx\nu_{\{j\}}$.
Now
\[
\sum_{|u|=1}\gamma_u\sigma^2_u
=\sum_{|u|=1}\nu_u\sigma^2_u[1+O(\eta+\eps)],
\]
where the constant in $O(\cdot)$ can depend
on $r$, and
\[
\sum_{|u|>1}\gamma_u\sigma^2_u
=\sum_{|u|>1}\nu_u\sigma^2_u\bigl[2^{|u|}+O(\eta+\eps)\bigr]
=O(\eta)\sum_{|u|=1}\nu_u\sigma^2_u.
\]
Similarly, $\sum_{|u|>1}\gamma_u\sigma^2_u
=O(\eta)\sum_{|u|=1}\nu_u\sigma^2_u$.
Therefore,
\[
\frac{\e_{\re}( \wt\var_{\bpb}(\bar X^*))}{\var
(\bar X)}
= \frac
{(1+O(\eta+\eps))\sum_{|u|=1}\nu_u\sigma_u^2}
{(1+O(\eta))\sum_{|u|=1}\nu_u\sigma_u^2}
= 1+O(\eta+\eps). 
\]
\upqed
\end{pf}

\subsection*{Proofs of Theorems \protect\ref{thmreffvarhet}
through \protect\ref{thmintboundshet}}

Here we prove the theorems for
the heteroscedastic case. We begin
with a lemma.
%
%
\begin{lemma}\label{lemereyyhet}
Let $\xbsi$ follow the heteroscedastic
random effects model (\ref{eqreffhet})
and let $\ybsi=\xbsi-\bar X$.
Then
%
%
\begin{equation}\label{eqexxhet}
\e_{\re}( \xbsi\xbsip) = \mu^2+\sneu\sigma^2_{\bsi,u}\oiuipu
\end{equation}
and
%
%
\begin{equation}\label{eqeyyhet}
\e_{\re}(\ybsi\ybsip)
=\sneu
(
\oiuipu\sigma^2_{\bsi,u}
-\nubsiu\sigma^2_{\bsi,u}-\nubsipu\sigma^2_{\bsip,u}
+\nusigbar).
\end{equation}
\end{lemma}
\begin{pf}
Equation (\ref{eqexxhet}) follows directly
just as the analogous expression did in Lemma~\ref{lemereyy}.
Once again, expanding
$\ybsi\ybsip$ yields
\[
\xbsi\xbsip
-\frac1N\sbsipp\zbsipp\xbsi\xbsipp
-\frac1N\sbsipp\zbsipp\xbsip\xbsipp
+\frac1{N^2}\sbsipp\sbsippp\zbsipp\zbsippp\xbsipp\xbsippp
\]
and we may assume that $\mu=0$ while
proving (\ref{eqeyyhet}).
Now
\[
\e_{\re}\biggl(\frac1N\sbsip\zbsip\xbsi\xbsip\biggr)
=\frac1{N}
\sneu\sbsip\zbsip\one_{\bsi_u=\bsip_u}\sigma_{\bsi,u}^2
=\sneu\sbsi\sigma_{\bsi,u}^2\nubsiu
\]
and
\begin{eqnarray*}
\e_{\re}\biggl(\frac1{N^2}\sbsipp\sbsippp\zbsipp\zbsippp\xbsipp
\xbsippp
\biggr)
&=&\frac1{N^2}\sneu
\sbsipp\sbsippp\zbsipp\zbsippp
\one_{i''_u=i'''_u}\sigma^2_{\bsi'',u}\\
&=&\frac1N\sneu\sbsipp\zbsipp\sigma_{\bsipp,u}^2\nubsippu\\
&=&\sneu\nusigbar,
\end{eqnarray*}
which together establish (\ref{eqeyyhet}).
\end{pf}
\begin{theorem}
In the heteroscedastic random effect model (\ref{eqreffhet})
%
%
\begin{equation}
\var( \bar X ) = \frac1N\sneu\sbsi\nubsiu\sigma^2_{\bsi,u}.
\end{equation}
\end{theorem}
\begin{pf}
The proof is very similar to that of Theorem~\ref{thmreffvar}.
\end{pf}
\begin{theorem}
In the heteroscedastic random effects model (\ref{eqreffhet})
\setcounter{equation}{19}
\begin{equation}
\e_{\re}( \wt\var_{\bpb}(\bar X^*))
= \frac1N\sneu\sbsi\gamma_{\bsi,u}\sigma^2_{\bsi,u},%
\end{equation}
where
\begin{equation}
\gamma_{\bsi,u}
=\sum_{k=0}^r(1+\tau^2)^k
(\nu_{\bsi,k,u} -2\nu_{\bsi,k}\nu_{\bsi,u}+\nukbar\nu_{\bsi,u}).
\end{equation}
\end{theorem}
\begin{pf}
We begin along the same lines as Theorem~\ref{thmnwvar}
and find that
\[
\e_{\re}(\wt\var_{\bpb}(\bar X^*))
=
\frac1{N^2}\sbsi\sbsip\zbsi\zbsip\e_{\re}(\ybsi\ybsip)\e
_{\bpb}(\wbsi
\wbsip).
\]
As in Theorem~\ref{thmpwvar},
$\e_{\bpb}(\wbsi\wbsip) 
=(1+\tau^2)^{|\miip|}$.

From Lemma~\ref{lemereyyhet},
%
%
\setcounter{equation}{29}
\begin{eqnarray}\label{eqhetparts}
&&\e_{\re}(\wt\var_{\bpb}(\bar X^*))\nonumber\\
&&\qquad=
\frac1{N^2}
\sum_{u\ne\varnothing}
\sbsi\sbsip\zbsi\zbsip(1+\tau^2)^{|\miip|}\\
&&\qquad\quad\hspace*{65.3pt}{}\times(
\one_{\bsi_u=\bsip_u} \sigma^2_{\bsi,u}
-\nubsiu\sigma^2_{\bsi,u}
-\nubsipu\sigma^2_{\bsip,u}
+\nusigbar).\nonumber
\end{eqnarray}
The contribution from the last term
in the parentheses of (\ref{eqhetparts}) is
\[
\frac1{N}\sneu\nusigbar\sum_{k=0}^r(1+\tau^2)^k\sbsi\zbsi\nu
_{\bsi,k}
=\sneu\nusigbar\sum_{k=0}^r(1+\tau^2)^k\nukbar.
\]

Therefore, the coefficient of $\sigma^2_{\bsi,u}$,
in $\e_{\re}(\wt\var_{\bpb}(\bar X^*))$
(when $\zbsi=1$) is
\begin{eqnarray*}
&&\frac1{N^2}\sbsip\zbsip\sum_{k=0}^r
\one_{|\miip|=k}(1+\tau^2)^k
( \one_{\bsi_u=\bsip_u} - 2\nubsiu)
+\frac{\nu_{\bsi,u}}N\sum_{k=0}^r(1+\tau^2)^r\nukbar\\
&&\qquad=
\frac1N
\sum_{k=0}^r(1+\tau^2)^k
(\nu_{\bsi,k,u} -2\nu_{\bsi,k}\nu_{\bsi,u}+\nukbar\nu_{\bsi,u}).
\end{eqnarray*}
\upqed
\end{pf}
\begin{theorem}
In the heteroscedastic random effects model (\ref{eqreffhet}),
the gain coefficient $\gamma_{\bsi,u}$
of (\ref{eqdefgainhet})
for $\zbsi=1$ and $u\ne\varnothing$
in the product reweighted bootstrap is
\[
\gamma_{\bsi,u} = \nu_{\bsi,u}\bigl[ (1+\tau^2)^{|u|}-1
+\theta_u\err\bigr] + \sum_{v\supsetneq u}(1+\tau^2)^{|v|}(\tau
^2)^{|v-u|}\nu
_{\bsi,v},
\]
where
$|\theta_u|\le(1+\tau^2)((1+\tau^2)^r-1)/\tau^2$.
For $\tau^2=1$
\[
\gamma_{\bsi,u} = \nu_{\bsi,u}\bigl[ 2^{|u|}-1
+\theta_u\err\bigr] + \sum_{v\supsetneq u}2^{|v|}\nu_{\bsi,v},
\]
where
$|\theta_u|\le2^{r+1}-2$.
\end{theorem}
\begin{pf}
From Theorem~\ref{thmpwvarhet},
$\gamma_{\bsi,u}
=\sum_{k=0}^r(1+\tau^2)^k
(\nu_{\bsi,k,u} -2\nu_{\bsi,k}\nu_{\bsi,u}+\nukbar\nu_{\bsi,u})$.
The proof is similar to that of
Theorem~\ref{thmbpbgains}, so we summarize
the steps. First,
\[
\sum_{k=0}^r(1+\tau^2)^k\nu_{\bsi,k,u}
= \sum_{v\supseteq u}\nu_{\bsi,v}(1+\tau^2)^{|v|}(\tau^2)^{|v-u|}.
\]
Next,
$\nu_{\bsi,0}=1-r\theta_{\bsi,0}\eps$ and
$\overline{\nu}_0 = 1-r\theta_{0}$,
while for $k\ge1$,
$\nu_{\bsi,k} = \theta_{\bsi,k}\eps$
and $\overline{\nu}_k = \theta_{k}\eps$.
Here, all of the $\theta$'s are in the interval $[0,1]$.
The result follows as in Theorem~\ref{thmbpbgains}.
\end{pf}
\begin{theorem}
For the heteroscedastic random effects model (\ref{eqreffhet}),
assume that there exist $m$ and $M$ with
$0<m\le\sigma^2_{\bsi,u}\le M<\infty$.
Then the product reweighted bootstrap with $\tau^2=1$
satisfies
\[
\frac{\e_{\re}( \wt\var_{\bpb}(\bar X^*))}{\var
(\bar X)}
= 1+O(\eta+\eps).
\]
\end{theorem}
\begin{pf}
First we show that main effects
dominate.
For $|u|>1$,
\begin{eqnarray*}
\sbsi\gamma_{\bsi,u}\sigma^2_{\bsi,u}
&\le& M\sbsi\nu_{\bsi,u}\bigl(2^{|u|}-1+2^{r+1}\eps\bigr)
+\sum_{v\supsetneq u}2^{|v|}\nu_{\bsi,v}\\
& = & M\biggl(
\nu_u\bigl(2^{|u|}-1+2^{r+1}\eps\bigr)
+\sum_{v\supsetneq u}2^{|v|}\nu_{v}\biggr)\\
& = &\bigl(2^{|u|}-1\bigr)M\nu_u\bigl(1+O(\eps+\eta)\bigr)\\
& = &O(\eta)\max_{1\le j\le r}\nu_{\{j\}}
\end{eqnarray*}
and, similarly, $\sbsi\nu_{\bsi,u}\sigma^2_{\bsi,u} =
O(\eta)\max_{1\le j\le r}\nu_{\{j\}}$.
For $u=\{j\}$,
\begin{eqnarray*}
\sbsi\gamma_{\bsi,\{j\}}\sigma^2_{\bsi,\{j\}}
& \ge& m\sbsi\nu_{\bsi,\{j\}}(1-2^{r+1}\eps)\\
& = & m\nu_{\{j\}}\bigl(1+O(\eps)\bigr).
\end{eqnarray*}
Therefore,
\[
\frac{\e_{\re}( \wt\var_{\bpb}(\bar X^*))}{\var
(\bar X)}
=\frac{\sbsi\sum_{j=1}^r\gamma_{\bsi,\{j\}}\sigma^2_{\bsi,\{j\}}}
{\sbsi\sum_{j=1}^r\nu_{\bsi,\{j\}}\sigma^2_{\bsi,\{j\}}}\bigl(1+O(\eta
+\eps)\bigr).
\]

Next we show that the main effects are
properly estimated
\begin{eqnarray*}
{\sbsi\sum_{j=1}^r\bigl|\gamma_{\bsi,\{j\}}-\nu_{\bsi,\{j\}}\bigr|\sigma
^2_{\bsi
,\{j\}}}
&\le& M{\sbsi\sum_{j=1}^r\bigl|\gamma_{\bsi,\{j\}}-\nu_{\bsi,\{j\}}\bigr|}\\
&\le&
M\sbsi\sum_{j=1}^r\nu_{\bsi,\{j\}}(2^{r+1}\eps+3^r\eta)\\
& = &\sum_{j=1}^r\nu_{\{j\}}O(\eta+\eps),
\end{eqnarray*}
while
$\sbsi\sum_{j=1}^r\nu_{\bsi,\{j\}}\sigma^2_{\bsi,\{j\}}
\ge m\sum_{j=1}^r\nu_{\{j\}}$.
\end{pf}
\end{appendix}

\section*{Acknowledgments}

We thank Paul Jones and Jonathan Chang for their assistance. We thank
Omkar Muralidharan and the reviewers for helpful comments.


%

\printaddresses

\end{document}